%% file: main.tex
\documentclass[sigplan,screen]{acmart}
\usepackage[]{hyperref}
\usepackage{url}
\usepackage{multirow} 
\usepackage{listings}
\usepackage{dblfloatfix}
\usepackage{placeins}  

\copyrightyear{2025}
\acmYear{2025}
\setcopyright{cc}
\setcctype{by-nc-sa}
\acmConference[ASPLOS '25]{Proceedings of the 30th ACM International Conference on Architectural Support for Programming Languages and Operating Systems, Volume 2}{March 30-April 3, 2025}{Rotterdam, Netherlands}
\acmBooktitle{Proceedings of the 30th ACM International Conference on Architectural Support for Programming Languages and Operating Systems, Volume 2 (ASPLOS '25), March 30-April 3, 2025, Rotterdam, Netherlands}
\acmDOI{10.1145/3676641.3716267}
\acmISBN{979-8-4007-1079-7/25/03}

\begin{document}

\title{PIM Is All You Need: A CXL-Enabled GPU-Free System for Large Language Model Inference}

%

\author{Yufeng Gu}
\authornote{Yufeng Gu and Alireza Khadem contributed equally to this research}
\affiliation{%
  \institution{University of Michigan}
  \city{Ann Arbor}
  \country{USA}}
\email{yufenggu@umich.edu}

\author{Alireza Khadem}
\authornotemark[1]
\affiliation{%
  \institution{University of Michigan}
  \city{Ann Arbor}
  \country{USA}}
\email{arkhadem@umich.edu}

\author{Sumanth Umesh}
\affiliation{%
  \institution{University of Michigan}
  \city{Ann Arbor}
  \country{USA}}
\email{sumanthu@umich.edu}

\author{Ning Liang}
\affiliation{%
  \institution{University of Michigan}
  \city{Ann Arbor}
  \country{USA}}
\email{nliang@umich.edu}

\author{Xavier Servot}
\affiliation{%
  \institution{ETH Zürich}
  \city{Zürich}
  \country{Switzerland}}
\email{xservot@student.ethz.ch}

\author{Onur Mutlu}
\affiliation{%
  \institution{ETH Zürich}
  \city{Zürich}
  \country{Switzerland}}
\email{omutlu@gmail.com}

\author{Ravi Iyer}
\authornote{This research was done while the author was at Intel Corporation}
\affiliation{%
  \institution{Google}
  \city{Mountain View}
  \country{USA}}
\email{raviiyer20@gmail.com}

\author{Reetuparna Das}
\affiliation{%
  \institution{University of Michigan}
  \city{Ann Arbor}
  \country{USA}}
\email{reetudas@umich.edu}





\renewcommand{\shortauthors}{Yufeng Gu and Alireza Khadem et al.}

\definecolor{myblue}{RGB}{0, 0, 150}

\newcommand{\att}[0]{CENT}
\newcommand{\rf}[1]{Shared Buffer}
\newcommand{\Sota}{State-of-the-art}
\newcommand{\sota}{state-of-the-art}
\newcommand{\ali}[1]{\noindent{\textcolor{orange}{\bf \fbox{AK} {\it#1}}}}
\newcommand{\reetu}[1]{\noindent{\textcolor{blue}{\bf \fbox{RD} {\it#1}}}}
\newcommand{\yufeng}[1]{\noindent{\textcolor{purple}{\bf \fbox{YG} {\it#1}}}}
\newcommand{\sumanth}[1]{\noindent{\textcolor{red}{\bf \fbox{SU} {\it#1}}}}
\newcommand{\ning}[1]{\noindent{\textcolor{cyan}{\bf \fbox{NL} {\it#1}}}}
\newcommand{\todo}[1]{\noindent{\textcolor{cyan}{\bf \fbox{TODO} {\it#1}}}}
\newcommand{\red}[1]{\textcolor{myblue}{#1}}
\newcommand{\ignore}[1]{}

\input{0_abstract}



\received{24 June 2024}
\received[revised]{2 October 2024}
\received[accepted]{27 January 2025}

\maketitle


\input{1_introduction}
\input{2_motivation}
\input{2_background}
\input{3_architecture}

\input{4_model_mapping}
\input{5_methodology}

\input{6_results}

\input{7_related_work}

\input{8_conclusion}

\begin{acks}
We thank the anonymous reviewers for their valuable feedback. This work was generously supported by NSF CAREER-1652294, NSF-1908601 and Intel gift awards. SAFARI authors acknowledge support from the Semiconductor Research Corporation, ETH Future Computing Laboratory (EFCL), AI Chip Center for Emerging Smart Systems Limited (ACCESS), and the European Union’s Horizon Programme for research and innovation under Grant Agreement No. 101047160.
\end{acks}


\appendix

\input{9_artifact}

\bibliographystyle{plainurl}
\bibliography{references}

\end{document}

%% file: 0_abstract.tex
\begin{abstract}

Large Language Model (LLM) inference uses an autoregressive manner to generate one token at a time, which exhibits notably lower operational intensity compared to earlier Machine Learning (ML) models such as encoder-only transformers and Convolutional Neural Networks.
At the same time, LLMs possess large parameter sizes and use key-value caches to store context information.
Modern LLMs support context windows with up to 1 million tokens to generate versatile text, audio, and video content.
A large key-value cache unique to each prompt requires a large memory capacity, limiting the inference batch size.
Both low operational intensity and limited batch size necessitate a high memory bandwidth.
However, contemporary hardware systems for ML model deployment, such as GPUs and TPUs, are primarily optimized for compute throughput. 
This mismatch challenges the efficient deployment of advanced LLMs and makes users pay for expensive compute resources that are poorly utilized for the memory-bound LLM inference tasks.

We propose \att{}, a \underline{C}XL-\underline{EN}abled GPU-Free sys\underline{T}em for LLM inference, which harnesses CXL memory expansion capabilities to accommodate substantial LLM sizes, and utilizes near-bank processing units to deliver high memory bandwidth, eliminating the need for expensive GPUs.
\att{} exploits a scalable CXL network to support peer-to-peer and collective communication primitives across CXL devices.
We implement various parallelism strategies to distribute LLMs across these devices.
Compared to GPU baselines with maximum supported batch sizes and similar average power, \att{} achieves 2.3$\times$ higher throughput and consumes 2.9$\times$ less energy.
\att{} enhances the Total Cost of Ownership (TCO), generating 5.2$\times$ more tokens per dollar than GPUs.

\end{abstract}

%% file: 1_introduction.tex
\section{Introduction}

Generative Artificial Intelligence (GenAI) has become pivotal in transforming a myriad of sectors. In the realm of content creation, Large Language Models (LLMs)~\cite{openai2023gpt4, claude, gemini-pro, touvron2023llama} provide assistance in writing, summarizing, and translating across diverse languages, revolutionizing the way textual content is produced.
LLMs are reshaping various fields in daily life, such as generating creative arts~\cite{dalle3, sora}, customer services through chatbots, generating code and debugging assistance in software development~\cite{kasneci2023chatgpt}. 
However, harnessing the power of LLMs presents substantial economic challenges, underlined by their significant resource requirements. A business cost model indicates that running ChatGPT inference tasks requires $\sim$3617 HGX A100~\cite{hgxa100} servers and costs $\sim$\$694,444 per day~\cite{chatgpt-cost}. 
Therefore, efficient and cost-effective server farms play a critical role in the broader adoption and practical application of LLMs.

Decoder-only LLMs have witnessed exponentially larger parameter sizes. 
At the same time, LLMs use key-value (KV) caches to store context information, and modern LLMs support context windows from 128K to 1M to generate versatile texts, audios, and videos~\cite{gpt4-turbo, gemini-pro}.
Both the model parameters and KV caches require a large memory capacity. To meet this demand, advanced GPU stations feature multiple GPUs. However, the computational resources of multi-GPU systems are often underutilized in LLM inference tasks. Unlike earlier ML models, LLMs exhibit lower operational intensity characteristics, necessitating high memory bandwidth, primarily due to the sequential token generation and the lack of inherent parameter reuse. Although batching strategies could mitigate this issue, KV caches specific to each user require large memory capacity, limiting the feasibility of high batch sizes.
Hence, the expensive compute throughput of GPUs and custom ML accelerators is significantly under-utilized for LLM inference because of the limited external memory bandwidth. 
As a result, \emph{users pay for expensive computing resources for memory-bound LLM inference tasks.}

\label{PIM}
The high cost and low compute utilization of GPU systems motivate an alternative solution for LLM inference tasks. Processing-In-Memory (PIM) architectures~\cite{aim1, aim2, aim3, aim4, fimdram, upmem, axdimm, to-pim, floatpim, sky, ultra, impala, aquabolt, sparsep, binary} place processing units (PU) adjacent to DRAM banks within memory chips, facilitating a significantly higher internal bandwidth.
However, near-bank PUs, fabricated in the DRAM process, impose a high area overhead that reduces the memory density.
A lower memory density is especially detrimental to LLMs with large memory requirements.
On the other hand, Processing-Near-Memory (PNM) architectures~\cite{cxl-pnm, samsung_pimpnm, noise, recnmp, enmc, HB-PNM, nda, trim, gomez2023evaluating, oliveira2022accelerating, gomez2022benchmarking} employ compute units near memory chips, \textit{e.g.,} in memory controllers.
PNM units are manufactured using CMOS process, offering more area-efficient compute capability at the cost of lower memory bandwidth compared to PIM.

To address these challenges, \att{} exploits Compute eXpress Link (CXL)~\cite{CXL} based memory expansion to provide the requisite memory capacity for LLMs.
\att{} establishes a practical CXL network to interconnect CXL devices.
Each CXL device consists of 16 memory chips, with each chip containing two GDDR6-PIM channels, and compute units near these memory chips (PNM).
This hierarchical PIM-PNM design supports the entire transformer block computation, eliminating the need for expensive GPUs.

\att{} uses a CXL switch to connect multiple CXL devices, that are driven by a host CPU.
The \textit{inter-device communication} is enabled by CXL transactions~\cite{CXL}.
The \textit{intra-device communication} between PIM chips and PNM units is supported through a \rf{}.
Using these protocols, \att{} provides peer-to-peer and collective communication primitives such as \textit{send/receive}, \textit{broadcast}, \textit{multicast} and \textit{gather}.
These primitives enable various parallelism strategies, efficiently distributing LLMs across CXL devices.
In \textit{Pipeline Parallel (PP)}~\cite{gpipe} mapping, we assign each transformer block to multiple memory channels within a single CXL device, facilitating the concurrent processing of multiple prompts on different pipeline stages.
PP prioritizes inference throughput to accommodate a large user base.
In \textit{Tensor Parallel (TP)}~\cite{alpa, megatron} mapping, we distribute a transformer block across all CXL devices.
TP focuses on reducing latency for real-time applications, providing smooth user experiences~\cite{fowers2018configurable}. 
We also explore hybrid TP-PP mappings to strike a balance between the latency and throughput.

Within a CXL device, we introduce the detailed mapping of a transformer block onto the hierachical PIM-PNM architecture. 
In PIM chips, near-bank PUs incorporate Multiply-Accumulate (MAC) units, which support more than 99\% of the arithmetic operations within a transformer block. 
The PNM units are composed of accelerators and RISC-V cores to perform other special and complex operations, such as Softmax, square root, and division. The integration of RISC-V cores allows for the flexible support of a wide range of LLMs.

In summary, this paper makes the following contributions:
\begin{itemize}
\item We propose \att{}, a GPU-free system that uses CXL memory expansion to accommodate the considerable memory capacity requirements of LLMs. We design a hierarchical PIM-PNM architecture to support the entire transformer block computation, eliminating the need for expensive GPUs.

\item We introduce a scalable CXL network to support collective and peer-to-peer communication primitives.
We describe the mapping of LLM parallelization strategies across CXL devices based on the CXL communication primitives.

\item We evaluate \att{} on Llama2~\cite{touvron2023llama} models. 
Compared to state-of-the-art GPUs with maximum supported batch sizes and similar average power, \att{} achieves 2.3$\times$ higher throughput and consumes 2.9$\times$ less energy. 
\att{} exhibits a lower Total Cost of Ownership (TCO), generating 5.2$\times$ more tokens per dollar than GPUs\footnote{Open-source \att{} simulator \href{https://github.com/Yufeng98/CENT/}{https://github.com/Yufeng98/CENT/}}.
\end{itemize}

\att{} is evaluated on Llama2 70B with a context length of up to 32K, but it can show higher benefits for larger model sizes and extended context lengths. As model sizes scale up, such as Grok 314B~\cite{grok}, Llama3 405B~\cite{llama3}, and DeepSeek-V3 671B~\cite{deepseek-v3}, inference serving demands significantly more hardware resources. In such cases, \att{} offers greater cost-efficiency compared to GPUs.

In reasoning tasks~\cite{O1-reasoning, deepseek-r1} and video generation~\cite{sora, polyak2024movie, gemini-pro}, where context length can range from tens of thousands to 1 million tokens, \att{} achieves higher throughput speedup due to its high memory bandwidth, which enhances memory-bound attention computations. Notably, GPUs can still benefit from long-text and video understanding tasks, as the prefill stage exhibits high operational intensity. In these scenarios, prefill and decoding processes can be disaggregated between GPUs and \att{}, respectively~\cite{zhong2024distserve, patel2024splitwise}.

%% file: 2_motivation.tex
\section{Motivation}~\label{sec:motivation}


The exponential growth of LLM parameters requires multi-GPU systems to accommodate the requisite memory capacity. However, LLMs exhibit limited operational intensity, making them memory-bound and resulting in suboptimal GPU utilization. Consequently, LLM service providers are paying significant costs for substantial computational throughput of multiple GPUs, which remains largely under-utilized.

\textbf{High Memory Capacity Requirement.} LLM parameter size has witnessed an exponential increase from Billion to Trillion magnitudes, far surpassing previous Machine Learning (ML) models.
In addition, the context windows that modern LLMs support range from 128K to 1M~\cite{gpt4-turbo, gemini-pro}, enabling them to understand and generate longer contents.
The long context window results in large KV caches, requiring substantial memory capacity.
These KV caches are unique to each user, further limiting the ability to scale up the inference batch size due to the memory capacity requirement.

\textbf{Low Operational Intensity.} LLM inference has two stages:
(a) The \textit{prefill} stage concurrently encodes input tokens within a prompt using matrix-matrix multiply (GEMM) operations.
(b) The \textit{decoding} stage decodes output tokens sequentially with matrix-vector multiply (GEMV) operations.
The operational intensity of GEMV is substantially lower than GEMM.
To mitigate this, several techniques are applied.
Batching strategies combine GEMV operations across multiple queries of a batch into GEMM operations.
This technique improves the operational intensity non-linearly because attention calculations rely on unique KV caches of each prompt.
Grouped-query attention~\cite{gqa} merges multiple GEMV operations into narrow GEMM, but its operational intensity still remains less than the GPU capabilities.

\begin{figure}[h]
\centering
    \includegraphics[width=\columnwidth]{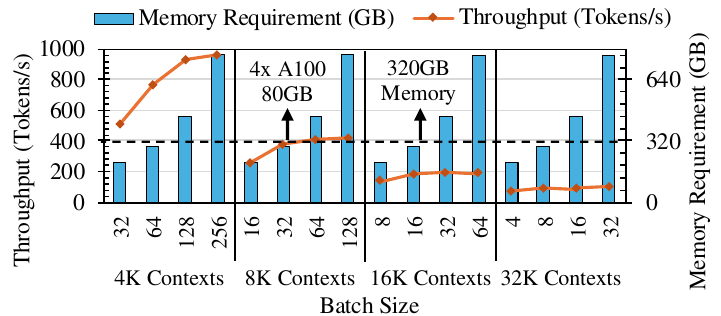}
    \caption{Llama2-70B~\cite{touvron2023llama, longlora} inference throughput and memory requirement on 4 A100 80GB GPUs.}
    \label{fig:Context_Length}
\end{figure}

\begin{figure}[h]
\centering
    \includegraphics[width=8cm]{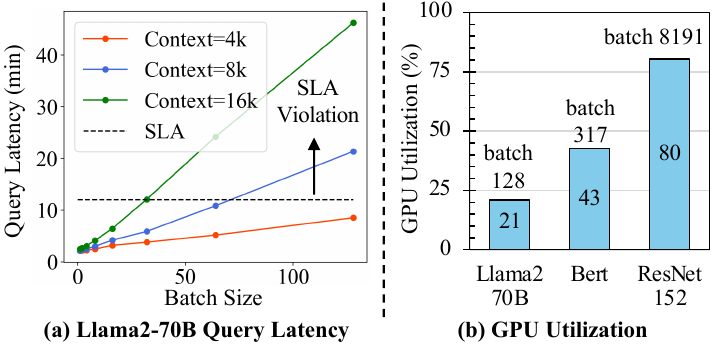}
    \caption{(a) Llama2-70B inference query latency increases with larger batches on 4 A100 80GB GPUs,  Prompt size=512, Decoding size=3584. (b) GPU compute utilization, measured by Nvidia Nsight Compute profiler on 4 GPUs for Llama2-70B and 1 GPU for the other two models.}
    \label{fig:GPU_utilization}
\end{figure}

\label{GPU Performance Characterization}
\textbf{GPU Performance Characterization.} We use vLLM~\cite{vLLM}, the \sota{} inference serving framework, to study the effect of batch size and context length on 4 Nvidia A100 80GB GPUs running the Llama2-70B model~\cite{touvron2023llama, longlora}.
Figure~\ref{fig:Context_Length} shows that inference throughput improves with larger batch sizes but reaches a plateau once the memory requirement exceeds the GPU memory size.
As context length increases, inference throughput saturates with even smaller batch sizes, from batch=128 at 4K context length to batch=8 at 32K context length.
Moreover, Figure~\ref{fig:GPU_utilization}(a) shows that LLM inference query latency increases with larger batch sizes and longer contexts, violating a realistic query latency Service Level Agreements (SLA) constraint~\cite{mlperf-sla}.

Figure~\ref{fig:GPU_utilization}(b) compares the GPU compute utilization of an LLM (Llama2-70B~\cite{touvron2023llama, longlora}) with an encoder-only transformer model (BERT~\cite{devlin2018bert}) and a Convolutional Neural Network (ResNet-152~\cite{he2016deep}).
BERT and ResNet-152 models predominantly consist of GEMM operations with high operational intensity, effectively utilizing GPU compute throughput.
Conversely, LLama2-70B exhibits limited operational intensity, resulting in a mere 21\% utilization of the available GPU compute throughput.
Finally, decoding an output token in the decoding stage takes $3.4\times$ longer than encoding a prompt token in the prefill stage due to the significant lower operational intensity of GEMV operations.

\label{PIM-prototype}
\textbf{PIM Provides Higher Memory Bandwidth}. 
Table~\ref{tab:hardware_comparison} compares various manufactured industrial PIM prototypes and GPU. 
PIM enables the compute units to utilize the internal memory bandwidth, which significantly exceeds the external memory bandwidth of high-end GPUs with high bandwidth memories (HBM). For example, GDDR6-based AiM~\cite{aim1, aim2} achieves $16~TB/s$ internal memory bandwidth compared to $2~TB/s$ external bandwidth of an A100 GPU with five HBM2E memory stacks. This large internal bandwidth coupled with a lower operational intensity makes PIM architectures a suitable alternative for expensive GPUs to perform LLM inference tasks. 

\begin{table}[h]
    \footnotesize
    \centering
    \caption{Hardware System Comparison}
    \label{tab:hardware_comparison}    
    \begin{tabular}{|c|c|c|c|c|}
        \hline
        Type & \multicolumn{3}{|c|}{PIM} & GPU   \\
        \hline
        Name & UPMEM & AiM & FIMDRAM & A100 \\
        \hline
        \hline
        Mem. Units & 8 DIMMs & 32 channels & 5 stacks & 5 stacks \\
        \hline
        Ex. BW (TB/s) & 0.15 & 1 & 1.5 & 2 \\
        \hline
        In. BW (TB/s) & 1 & 16 & 12.3 & - \\
        \hline
        Capacity (GB) & 64 & 16 & 30 & 80 \\
        \hline
        TFLOPS & 0.5 TOPS\footnotemark & 16 & 6.2 & 312 \\
        \hline
        Ops/Byte & 0.5 & 1 & 0.5 & 156 \\
        \hline
        Mem. Density & 25\%-50\% & 75\% & 75\% & - \\
        \hline
    \end{tabular}
\end{table}

\textbf{Low Memory Density of PIM.}
PIM suffers from a lower memory density due to the near-bank processing units that are fabricated in the DRAM process.
For instance, DDR4-based UPMEM R-DIMM and GDDR6-based AiM reduce the memory capacity to $25\%{-}50\%$ and $75\%$ compared to conventional DDR4 R-DIMMs and GDDR6 channels, respectively~\cite{upmem, aim2}.
An HBM2-based FIMDRAM cube consists of 4 PIM-enabled DRAM dies with $50\%$ memory density and 4 conventional dies, lowering the memory capacity by $25\%$ on average~\cite{fimdram}.
Given the lower memory density of PIM technologies and the substantial memory demands of LLMs, leveraging PIM as a scalable solution for LLMs presents significant challenges.

\footnotetext{UPMEM supports only integer precision, so unit is TOPs.}

\textbf{Scalable Network of PIM.}
Scaling the memory capacity of PIM-enabled memories requires a scalable interconnect, efficient collective communication primitives, and parallelization strategies to optimally map LLMs to PIM devices.
We utilize CXL 3.0~\cite{CXL} as a low-latency interconnect protocol, built on top of the PCIe physical layer.
CXL 3.0 supports inter-device communication through a CXL switch.
Compared to network-based RDMA, CXL.mem offers ${\sim}8\times$ lower latency~\cite{gouk2022direct}.
The CXL 3.0 protocol can support up to 4,096 nodes, exhibiting better scalability than NVLink~\cite{nvlink}.
NVLink provides higher bandwidth (at a higher cost), which is critical for LLM training.
However, we show that the lower bandwidth of CXL is not a bottleneck for LLM inference due to the limited volume of data transfers in various parallelization strategies.

To distribute the LLMs, we detail the mapping of the transformer blocks to the CXL devices based on the Pipeline Parallel (PP)~\cite{megatron, alpa} and Tensor Parallel (TP)~\cite{gpipe} strategies.
For PP, we provide peer-to-peer \textit{send} and \textit{receive} primitives for the transmission of the embedding vector between the pipeline stages across CXL devices.
For TP, we implement \textit{gather} and \textit{broadcast} collective communication primitives to transfer partial results.
To balance the throughput and latency of the network, we study the hybrid TP-PP parallelization strategy using the \textit{multicast} primitive.

\textbf{Hierarchical PIM-PNM Architecture.}
In addition to GEMV, a transformer block contains different layers, such RMSNorm~\cite{rmsnorm-paper}, Rotary Embedding~\cite{rope-paper}, and SiLU~\cite{silu-paper}.
For the end-to-end execution of transformer blocks as an alternative to costly GPUs, there are two options:
(a) Perform all operations near-bank using a general-purpose PU similar to UPMEM~\cite{upmem} architecture.
(b) Perform MAC operations of GEMVs in domain-specific near-bank PUs similar to AiM~\cite{aim2}, and assign other operations to the PNM units, shared by multiple PIM chips.
We use the second approach and propose a hierarchical PIM-PNM solution because of two primary reasons:
First, a general-purpose near-bank PU incurs more overhead on memory density and yields lower compute throughput compared to domain-specific alternatives.
Second, MAC operations constitute over $99\%$ of arithmetic operations within a transformer block, rendering general-purpose near-bank PUs over-provisioned for other infrequent arithmetic operations.

%% file: 2_background.tex
\section{Background} \label{section:LLM}

\begin{figure}[t]
    \centering
    \includegraphics[width=8cm]{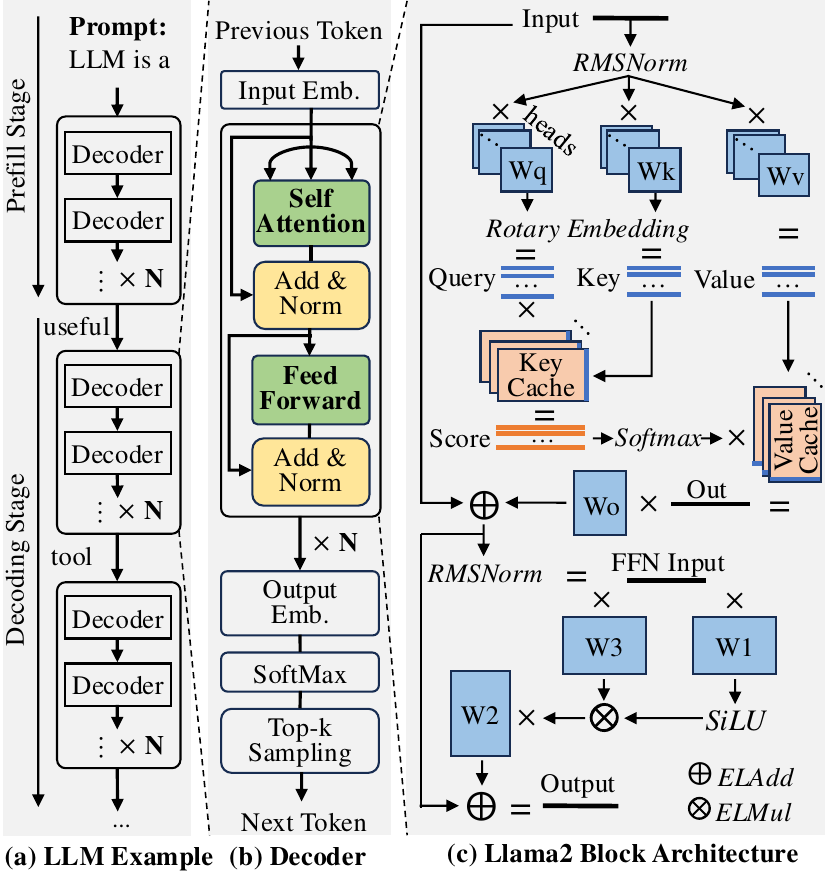}
    \caption{(a) Prefill stage encodes prompt tokens in parallel. Decoding stage generates output tokens sequentially.
    (b) LLM contains N$\times$ decoder transformer blocks. 
    (c) Llama2 model architecture.}
    \label{fig:LLaMA_model}
\end{figure}

Figure~\ref{fig:LLaMA_model}(a) shows that a decoder-only LLM initially processes a user prompt in the “prefill” stage and subsequently generates tokens sequentially during the “decoding” stage.
Both stages contain an input embedding layer, multiple decoder transformer blocks, an output embedding layer, and a sampling layer.
Figure~\ref{fig:LLaMA_model}(b) demonstrates that the decoder transformer blocks consist of a self attention and a feed-forward network (FFN) layer, each paired with residual connection and normalization layers. 





Figure~\ref{fig:LLaMA_model}(c) demonstrates the Llama2~\cite{touvron2023llama} model architecture as a representative LLM.
In the self-attention layer, query, key and value vectors are generated by multiplying input vector to corresponding weight matrices.
These matrices are segmented into multiple heads, representing different semantic dimensions.
The query and key vectors go though Rotary Positional Embedding (RoPE) to encode the relative positional information~\cite{rope-paper}.
Within each head, the generated key and value vectors are appended to their caches.
The query vector is multiplied by the key cache to produce a score vector.
After the Softmax operation, the score vector is multiplied by the value cache to yield the output vector.
The output vectors from all heads are concatenated and multiplied by output weight matrix, resulting in a vector that undergoes residual connection and Root Mean Square layer Normalization (RMSNorm)~\cite{rmsnorm-paper}.
The residual connection adds up the input and output vectors of a layer to avoid vanishing gradient~\cite{he2016deep}.
The FFN layer begins with two parallel fully connections, followed by a Sigmoid Linear Unit (SiLU), and ends with another fully connection.

%% file: 3_architecture.tex
\section{\att{} Architecture}

\begin{figure}[h]
	\centering
  	\includegraphics[width=7.5cm]{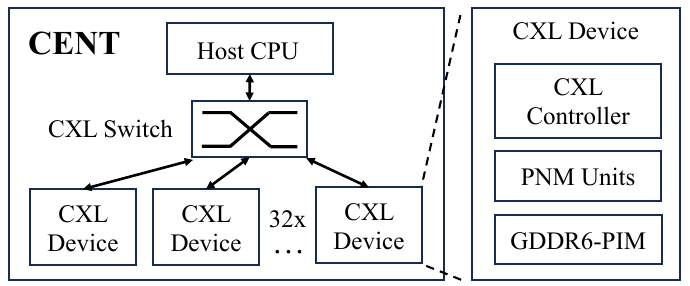}
    \caption{CENT Architecture.}
	\label{fig:Architecture_Top}
\end{figure}

Figure~\ref{fig:Architecture_Top} presents the \att{} architecture, where a CXL switch interconnects 32 CXL devices, driven by a host CPU. 
Each CXL device integrates a CXL controller, PNM units, and 16 memory chips, each equipped with two GDDR6-PIM channels (hereafter referred to as PIM channels).
We introduce a CXL-based network architecture, a hierarchical PIM-PNM design and the \att{} ISA in this section.

\subsection{CXL-based Network Architecture}
\label{sub:CXL-Network}

\att{} integrates the CXL 3.0 protocol, using the PCIe 6.0 physical interface. The CXL switch is connected to the host machine with \texttt{x16} lanes, whereas each CXL device is connected to the switch through \texttt{x4} lanes.
The switch supports the communication between the host and CXL devices, and peer-to-peer communication between CXL devices. 

\begin{figure}[h]
    \centering
    \includegraphics[width=\columnwidth]{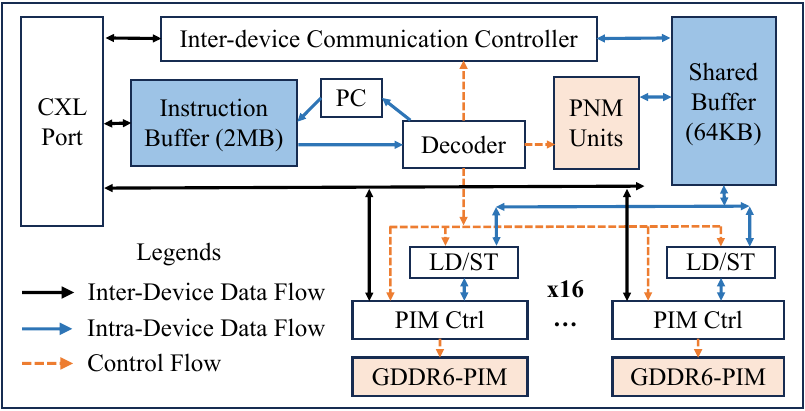}
    \caption{CXL Device Architecture.}
    \label{fig:CXL_device}
\end{figure}

\textbf{Inter-Device Communication.} 
Figure~\ref{fig:CXL_device} shows the architecture of a CXL device. Communication between CXL devices involves the \rf{} and is orchestrated by the inter-device communication controller in conjunction with the CXL port.
We introduce a \textit{broadcast} primitive, allowing one CXL device to write data to multiple devices through a single request. The standard CXL.mem protocol lacks this support. We implement it by using one of the reserved header codes within the Header slot (H-slot) of the Port Based Routing (PBR) flit. 
The H-slot is decoded by the switch for routing.
Upon identifying a flit encoded with this reserved H-slot code, the switch interprets it as a broadcast request and forwards the flit to designated CXL devices. We also modified the CXL port to (1) incorporate a device ID mask within the header slot of the broadcast message, and (2) expect write acknowledgements from all destination devices.

Inter-device communication is supported by \texttt{SEND\_CXL}, \texttt{RECV\_CXL} and \texttt{BCAST\_CXL} instructions. The non-blocking
\texttt{SEND\_CXL} specifies the device ID (\texttt{DVid}) and the \rf{} address in source and destination devices. Conversely, \texttt{RECV\_CXL} operates in a blocking manner and does \textit{not} specify a device ID. A pair of send and receive instructions constitutes a CXL write transaction. 
\texttt{BCAST\_CXL} is also non-blocking and uses an 8-bit \texttt{DVcount} parameter to specify the number of subsequent CXL devices to which the data is broadcast.
The \textit{multicast} primitive is supported in a similar manner.
To accomplish \textit{gather}, the receiving device executes multiple \texttt{CXL\_RECV} instructions, while each sender executes one \texttt{SEND\_CXL} instruction. Note that the receive instruction omits any device ID specification, thereby rendering the order of incoming CXL flits inconsequential.

\begin{figure}[h]
    \centering
    \includegraphics[width=0.40\textwidth]{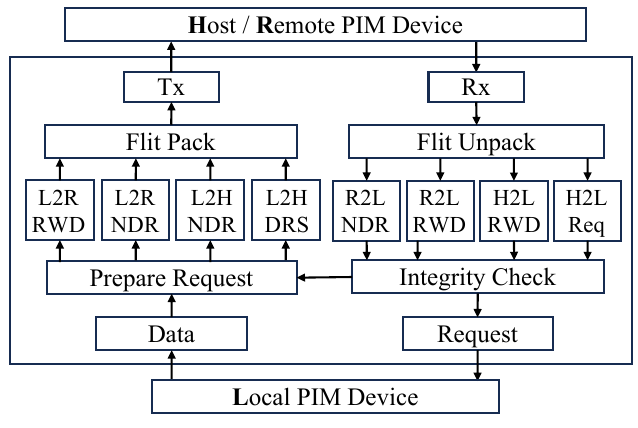}
    \caption{CXL Port Architecture.}
    \label{fig:CXL_port}
\end{figure}

\textbf{CXL Port} is depicted in Figure~\ref{fig:CXL_port}. 
CXL nodes are classified into three categories: Host (H), representing the host machine; Local (L), the CXL device we are considering; and Remote (R), referring to other CXL devices interconnected via the switch. CXL port is equipped with virtual channels. Requests from the host and remote nodes are unpacked onto the Rx H2L and R2L queues, and responses to the host and remote nodes are allocated to the Tx L2H and L2R queues.

Transactions comprise a request and a response. On the transmit (Tx) datapath, the CXL port packs requests into flits, which are unpacked on the receive (Rx) datapath by the destination device.
The CXL port supports 2 types of transactions: read transactions, initiated with a \textit{Request} (Req) and concluded with \textit{Data with Response} (DRS); and write transactions that begin with a \textit{Request with Data} (RWD) and finish with \textit{No Data Response} (NDR) acknowledgment.

\vspace{-3mm}

\subsection{Hierarchical PIM-PNM Architecture}~\label{subsec:pim_pnm_arch}

In Figure~\ref{fig:CXL_device}, \att{} instructions are transmitted from the host to a 2MB instruction buffer in each device. These instructions are further distributed to PIM channels and PNM units.
Standard read/write transactions are dispatched to PIM controllers similar to non-PIM memory modules. \att{} arithmetic instructions are decoded into micro-ops and subsequently directed to PIM controllers and PNM units. 

\begin{figure*}[htbp]
	\centering
  	\includegraphics[width=0.9\textwidth]{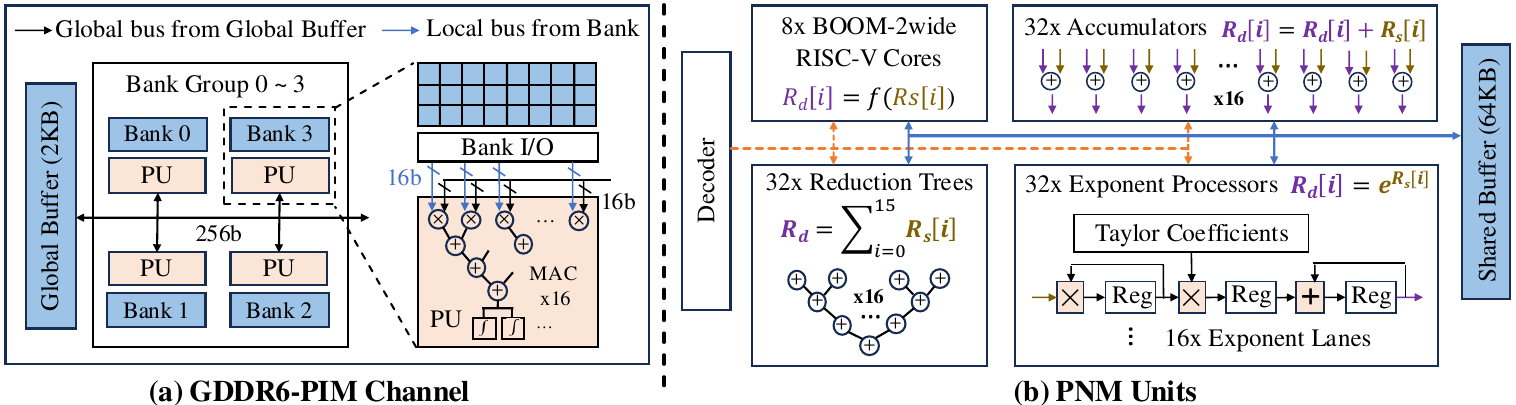}
    \caption{Hierarchical PIM-PNM Architecture}
	\label{fig:Architecture_PIM_PNM}
\end{figure*}

\textbf{GDDR6-PIM Channel.}
The CXL device integrates 16 PIM controllers, each managing two PIM channels. These controllers receive micro-ops from the decoder and convert them into DRAM commands.
Figure~\ref{fig:Architecture_PIM_PNM}(a) shows that the PIM channel consists of a \texttt{2KB} Global Buffer shared by four bank groups.
The bank group contains four banks.
Each bank has a 32MB memory capacity coupled with a near-bank PU.
%

Within the PU is a 16 MAC reduction tree, operating on Bfloat16 (\texttt{BF16}) data elements. Each multiplier receives 16-bit data directly from its associated local bank, in addition to another 16-bit data from either the Global Buffer or its neighboring bank (such as Bank $0$ and Bank $1$). 
The Global Buffer is capable of broadcasting 256-bit data to all PUs concurrently. 32 accumulation registers are incorporated to hold the MAC results in the PU and are designated by the \att{} ISA. The activation function (AF) leverages lookup tables stored within the DRAM bank and linear interpolation. 

The PU operates at 1GHz, equivalent to $t_{CCDS}$ ($2t_{CK}$) of the PIM bank, yielding a compute throughput of 32 GFLOPS.
PIM channels are optimized to allow 16 near-bank PUs to operate in parallel. To facilitate this, the PIM controller issues an activate-all-banks \texttt{ACTab} command, followed by PIM commands such as \texttt{MACab} and concludes with a precharge-all-banks \texttt{PREab} command. The \texttt{ACTab} command is enabled by the reservoir capacitors introduced in AiM~\cite{aim1, aim2}, and \texttt{PREab} is already supported by the GDDR6 DRAM~\cite{samsung-8gb-gddr6}.

\textbf{PNM Units.}
While near-bank PUs could efficiently support MAC operations, LLMs necessitate a broader set of operations beyond MACs. 
To address this, the CXL device incorporates the following PNM units, as shown in Figure~\ref{fig:Architecture_PIM_PNM}(b):
(1) \textit{32 Accumulators}: each retrieves two values from the \rf{} as inputs and segments the 256-bit inputs into 16 groups for \texttt{BF16} accumulations.
(2) \textit{32 Reduction Trees}: each fetches a single 256-bit value from the \rf{}, reducing 16 \texttt{BF16} input elements to a single \texttt{BF16} value. The result is stored into the first 16-bit element in a 256-bit \rf{} slot.
(3) \textit{32 Exponent Accelerators}, each accesses a 256-bit value from the \rf{}, dividing it into 16 lanes. In each lane, the exponent of a \texttt{BF16} input element is calculated by a 10-order Taylor Series approximation.
(4) \textit{8 BOOM-2wide RISC-V cores}~\cite{BOOM}, facilitating the execution of less common operations (such as square root and inversion), and accommodating future improvements in LLMs. Each RISC-V core is equipped with a \texttt{64KB} instruction buffer, which is initialized by the host through CXL write transactions.

\textbf{Intra-Device Communication} between PIM channels and PNM units is enabled through \att{} data movement instructions and a \texttt{64KB} \rf{}.
The \rf{} is viewed by PIM channels as 256-bit registers. 
\att{} facilitates data transfers between DRAM banks and the \rf{} by \texttt{WR\_SBK} and \texttt{RD\_SBK} instructions. These transfers are conducted by the load/store unit associated with each memory controller. Additionally, \texttt{WR\_ABK} instruction segments a 256-bit register into 16 discrete \texttt{BF16} values and concurrently stores them in the same row and column address of all 16 banks within a channel. Communication among banks in a PIM channel is mediated by the Global Buffer through \texttt{COPY\_BKGB} and \texttt{COPY\_GBBK} instructions. Similar to PIM channels, PNM units interface with the \rf{} at a 256-bit granularity and abstract it as a register file. The RISC-V core views the \rf{} as a byte-addressable memory and interacts with it through 16-bit loads and stores in a designated \texttt{64KB} region of the memory space.

\subsection{ISA Summary} \label{ISA_Summary}

Table~\ref{tab:ISA_ARITHMETIC} shows \att{} arithmetic instructions. The \texttt{CHmask} parameter directs the PIM decoder to broadcast micro-ops to specified PIM channels.
PIM decoder generates \texttt{OPsize} micro-ops from a single instruction, targeting subsequent \rf{} slots and DRAM column addresses.
The \texttt{Regid} parameter identifies the specific accumulation register within the PU, while \texttt{AFid} determines the type of non-linear activation function.
The \texttt{RISCV} instruction is designed to initiate the execution of RISC-V cores at the specific start program counter (PC) address.

\begin{table}[h]
    \footnotesize
    \centering
    \caption{\att{} Arithmetic Instructions}
    \label{tab:ISA_ARITHMETIC}
    \begin{tabular}{|c||c|}
        \hline
        \textbf{Instruction} & \textbf{Assembly} \\
        \hline
        \hline
        \multicolumn{2}{|c|}{\textbf{Near-Bank PUs}} \\
        \hline
        MAC All Bank & \texttt{MAC\_ABK CHmask OPsize RO CO Regid} \\
        \hline
        Element-wise Mult. & \texttt{EW\_MUL CHmask OPsize RO CO} \\
        \hline
        Activation Function & \texttt{AF CHmask AFid Regid} \\
        \hline
        \hline
        \multicolumn{2}{|c|}{\textbf{PNM Units}} \\
        \hline
        Exponent & \texttt{EXP OPsize Rd Rs} \\
        \hline
        Reduction & \texttt{RED OPsize Rd Rs} \\
        \hline
        Accumulation & \texttt{ACC OPsize Rd Rs} \\
        \hline
        RISCV operation & \texttt{RISCV OPsize PC Rd Rs} \\
        \hline
    \end{tabular}
\end{table}


Table~\ref{tab:ISA_MOVE} summarizes \att{} data movement instructions,
specifying DRAM bank locations using channel (\texttt{CHid}), bank (\texttt{BK}), row (\texttt{RO}), and column (\texttt{CO}).
The source and destination \rf{} addresses are specified by \texttt{Rd} and \texttt{Rs}.

\begin{table}[h]
    \footnotesize
    \centering
    \caption{\att{} Data Movement Instructions}
    \label{tab:ISA_MOVE}    
    \begin{tabular}{|c||c|}
        \hline
        \textbf{Instruction} & \textbf{Assembly} \\
        \hline
        \hline
        \multicolumn{2}{|c|}{\textbf{CXL Device $\leftrightarrow$ CXL Device}} \\
        \hline
        Send & \texttt{SEND\_CXL DVid Rs Rd} \\
        \hline
        Receive & \texttt{RECV\_CXL} \\
        \hline
        Broadcast & \texttt{BCAST\_CXL DVcount Rs Rd} \\
        \hline
        \multicolumn{2}{|c|}{\textbf{\rf{} $\leftrightarrow$ DRAM Banks}} \\
        \hline
        Write Single Bank & \texttt{WR\_SBK CHid OPsize BK RO CO Rs} \\
        \hline
        Read Single Bank & \texttt{RD\_SBK CHid OPsize BK RO CO Rd} \\
        \hline
        Write All Banks & \texttt{WR\_ABK CHid RO CO Rs Regid} \\
        \hline
        \multicolumn{2}{|c|}{\textbf{Global Buffer $\leftrightarrow$ DRAM Banks}} \\
        \hline
        Copy Bank $\rightarrow$ Global Buffer & \texttt{COPY\_BKGB CHmask OPsize RO CO} \\
        \hline
        Copy Global Buffer $\rightarrow$ Bank & \texttt{COPY\_GBBK CHmask OPsize RO CO} \\
        \hline
        \multicolumn{2}{|c|}{\textbf{\rf{} $\leftrightarrow$ PUs}} \\
        \hline
        Write bias & \texttt{WR\_BIAS CHmask Rs} \\
        \hline
        Read MAC register & \texttt{RD\_MAC CHmask Rd Regid} \\
        \hline
        \multicolumn{2}{|c|}{\textbf{\rf{} $\rightarrow$ Global Buffer}} \\
        \hline
        Write Global Buffer & \texttt{WR\_GB CHmask OPsize CO Rs} \\
        \hline
    \end{tabular}
\end{table}


%% file: 4_model_mapping.tex
\section{Model Mapping} \label{model mapping}

The ever-increasing parameter size of the LLMs, coupled with the lower memory density of PIM, necessitates the distribution of the LLM inference on a scalable network of PIM modules.
In this section, we introduce the mapping of various LLM parallelization strategies on \att{}'s CXL-based network architecture using the proposed collective and peer-to-peer communication primitives.

\subsection{Pipeline-Parallel Mapping (PP)}

\begin{figure}[h]
    \centering
    \includegraphics[width=8cm]{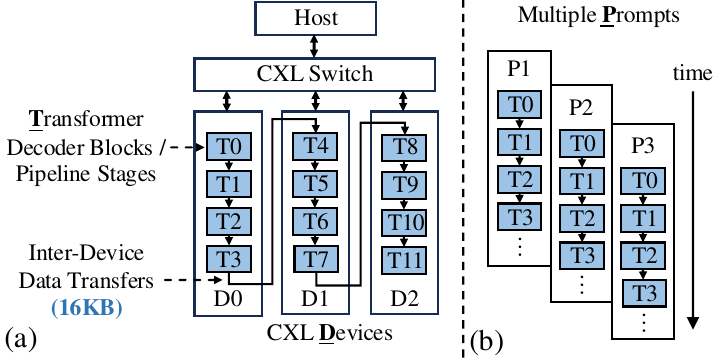}
    \caption{Pipeline parallelism: (a) Transformer decoder blocks are distributed across CXL devices and form the pipeline stages. Each block is mapped to multiple GDDR6-PIM channels. (b) Multiple prompts are executed in different stages of the pipeline.}
    \label{fig:Pipeline_Parallelism}
\end{figure}

Cloud providers serve a large user base, where inference throughput is crucial.
To improve throughput, PP~\cite{gpipe} assigns each transformer block to a pipeline stage.
The individual queries in a batch are simultaneously processed in different stages of the pipeline.
Figure~\ref{fig:Pipeline_Parallelism} shows that we map multiple pipeline stages (\textit{e.g.,} \texttt{T0-3}) to a CXL device (\textit{e.g.,} \texttt{D0}).
Each stage requires multiple PIM channels, depending on the memory requirements of the decoder block.
To prevent excessive communication and keep the latency of pipeline stages identical, we avoid splitting a pipeline stage between the PIM channels of two CXL devices.

In each iteration, the output of each transformer block is transferred to the next pipeline stage.
\att{} performs this data transfer using intra-device communication for pipeline stages within the same CXL device, and using peer-to-peer \textit{send} and \textit{receive} primitives for those in different CXL devices.
This CXL data transfer contains only an 8K embedding vector (\texttt{16KB} data) in Llama2-70B.
The CXL transfer latency of PP is negligible compared to PIM and PNM latencies.

Note that \att{} does not support batch processing within a single pipeline stage because of two primary reasons:
First, batching requires a significantly larger Global Buffer and \rf{} (Section~\ref{subsec:pim_pnm_arch}) to concurrently store the embedding vectors of multiple queries.
Second, batching enhances the operational intensity and compute utilization (Section~\ref{sec:motivation}), while PP fully utilizes PIM compute resources.
Therefore, applying batching on top of PP only increases the latency.

\subsection{Tensor-Parallel Mapping (TP)}

\begin{figure}[h]
    \centering
    \includegraphics[width=8cm]{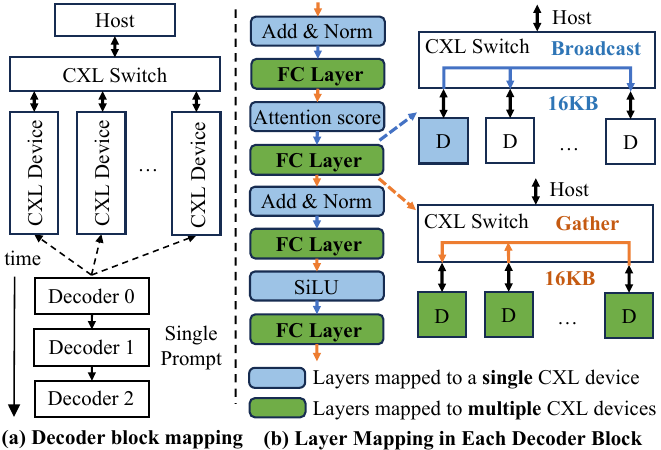}
    \caption{(a) Tensor parallelism: each transformer block is assigned to multiple CXL devices. Prompts are processed sequentially. (b) In a transformer block, fully connected layers are spread across CXL devices, while other operations are confined to a single device.}
	\label{fig:Model_Parallelism}
\end{figure}

Inference latency is critical in real-time applications to provide a smooth user experience~\cite{fowers2018configurable}.
To enhance the latency, TP~\cite{alpa, megatron} uses all compute resources to process decoder blocks one at a time.
To implement TP, Figure~\ref{fig:Model_Parallelism}(a) shows that \att{} assigns each transformer decoder block across all CXL devices.
Figure~\ref{fig:Model_Parallelism}(b) illustrates the detailed mapping of a transformer block using TP.
The infrequent residual connection and normalization layers are confined within a single master CXL device.
Distributing the attention layer requires the frequent use of expensive \textit{AllReduce} collective communication primitive, which significantly increases the CXL communication overhead~\cite{megatron}.
Consequently, the attention layer is mapped to the master CXL device.

Prior to the execution of an FC layer, the embedding vector (\texttt{16KB} for Llama2-70B) is \textit{broadcast} from the master CXL device to all devices via the CXL switch.
This enables each device to locally perform GEMV on multiple rows of the weight matrix.
Following the execution of an FC layer, partial result vectors are \textit{gathered} to the master CXL device.
This approach optimizes the execution of FC layers across multiple devices, while reducing the communication overhead of TP through the CXL switch to only \texttt{135KB} data transfer for each transformer block of the Llama2-70B model.

\subsection{Hybrid Tensor-Pipeline Parallel Mapping}~\label{subsec:hybrid_parallel}

The TP and PP mappings focus either on inference latency or throughput.
However, balancing both can be crucial in real-world deployment scenarios when considering Quality of Service (QoS) requirements~\cite{mlperf-sla}.
We explore a hybrid TP-PP strategy to achieve this balance, where each transformer decoder is allocated to multiple consecutive CXL devices. For example, among $32$ devices, mapping each decoder to $32/4=8$ devices enables TP=8 and PP=4.
The embedding vectors are \textit{multicast} and \textit{gathered} by the master CXL device of each pipeline stage.
This configuration effectively reduces token decoding latency by utilizing compute resources from multiple CXL devices (TP), while also improving the throughput by processing multiple prompts in parallel (PP).

\subsection{Transformer Block Mapping} \label{subsec:block_mapping}

\begin{figure}[h]
	\centering
    \includegraphics[width=\columnwidth]{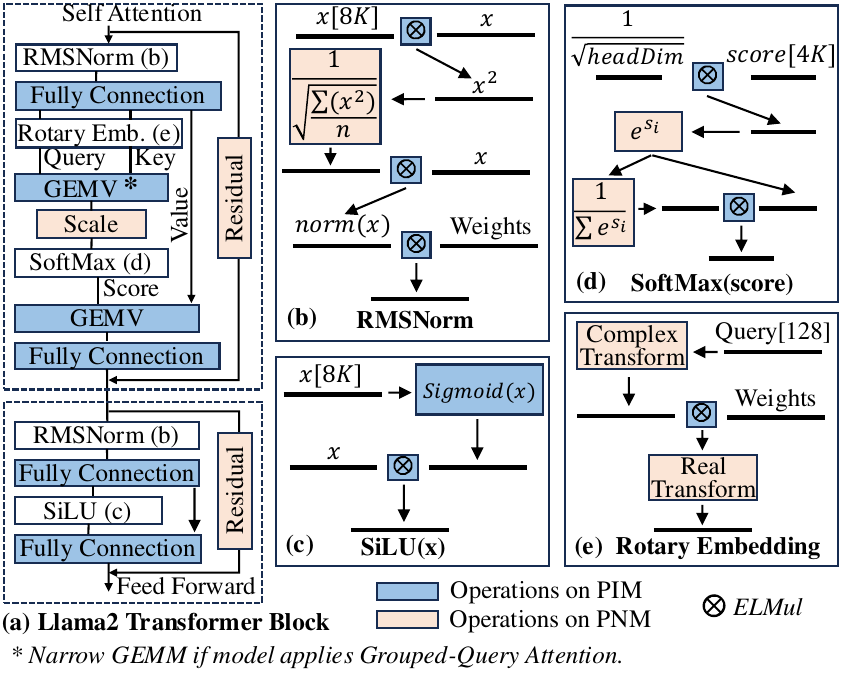}
    \caption{(a) Llama2-70B Transformer Block. Blue and orange operations are mapped to PIM and PNM PUs, respectively.
    (b)$\sim$(e) Operation mapping for RMSNorm, SiLU, SoftMax and Rotary embedding.}
	\label{fig:LLaMA_mapping}
\end{figure}

\att{} involves a fine-grained mapping of the transformer block onto CXL devices, PNM accelerators, and PIM channels.
This technique permits the complete execution of a transformer block within the CXL device, thereby eliminating the necessity for any interaction with the host system.
Figure~\ref{fig:LLaMA_mapping}(a) illustrates the operations within a Llama2 transformer block. 
Operations within the blue blocks are assigned to PIM channels, including GEMV in fully connected layers, vector dot product in RMSNorm, and element-wise multiplication in RMSNorm, SiLU, Softmax and Rotary Embedding, as detailed in Figure~\ref{fig:LLaMA_mapping}(b), (c), (d), and (e), respectively.
On the other hand, model-specific operations marked in orange, such as square root, division, Softmax, and vector addition in residual connections, are handled by the PNM's RISC-V cores and accelerators.
\att{} supports \textit{Grouped-Query Attention}~\cite{gqa} in Llama2-70B by unrolling GEMM to GEMV.

In Figure~\ref{fig:LLaMA_mapping}(d), the score dimension varies between $1$ and $4k$, accommodating the 4K sequence length in this example.
The embedding dimensions, as shown in Figure~\ref{fig:LLaMA_mapping}(b) and (c), are set to $8K$.
The rotary embedding process, depicted in Figure~\ref{fig:LLaMA_mapping}(e), begins with the RISC-V PNM cores transforming an attention head of dimension $128$ into $64$ groups of the complex number representations (\textit{e.g.,} $[a, b, c, d]$ to $[(a+jb), (c+jd)]$).
The PIM PUs within memory chips then multiply complex values and pre-loaded weights.
Finally, RISC-V PNM cores convert the computed results back to their real value representations.

\att{}'s PIM computations include three key operations.
This paragraph explains the execution of each operation within a GDDR6-PIM channel.
(a) \textit{GEMV}: The matrix is partitioned along its rows and distributed across all 16 banks. The vector is transferred to the Global Buffer. \texttt{MAC\_ABK} instructions then broadcast 256-bit vector segments from the Global Buffer to all near-bank PUs, retrieve 256-bit segments of the matrix rows from the banks, and perform MAC operations.
(b) \textit{Vector dot product}: In this operation, input vectors are stored in neighboring banks. \texttt{MAC\_ABK} instructions retrieve 256-bit segments from these banks and perform MAC operations. Throughout this process, only one of the two neighboring near-bank PUs is utilized.
(c) \textit{Element-wise multiplication}: Before this operation, input vectors are stored in two banks within each bank group, which consists of four banks. \texttt{EW\_MUL} instructions then retrieve 256-bit segments from these two banks, perform the multiplication, and store the results in another bank within the same bank group.

\subsection{End-to-End Model Mapping}~\label{subsec:e2e_model_mapping}

\att{} supports the end-to-end query execution in LLM inference tasks. In the prefill stage, \att{} processes tokens in the prompt one after another to fill out KV caches, using a similar approach to that in the decoding stage. 
Within each token, both input embeddings and transformer blocks are mapped to CXL devices using the mapping techniques introduced in Section~\ref{subsec:block_mapping}. In the decoding stage, after a series of transformer blocks, the top-k sampling operations are executed on the host CPU.

\subsection{Programming Model}

Users can specify the \att{} hardware configuration, including the number of PIM channels to utilize, and the number of pipeline stages. The tensor mapping strategy is determined by this configuration. \att{} library provides Python APIs to allocate memory space and load model parameters according to the model mapping strategy. 
These APIs also support commonly used LLM operations, such as \texttt{GEMV}, \texttt{LayerNorm}, \texttt{RMSNorm}, \texttt{RoPE}, \texttt{SoftMax}, \texttt{GeLU}, \texttt{SiLU}, \textit{etc.} 
\att{} uses an in-house compiler to generate arithmetic and data movement instructions illustrated in Section~\ref{ISA_Summary}.

\begin{figure}[h]
    \centering
    \includegraphics[width=8cm]{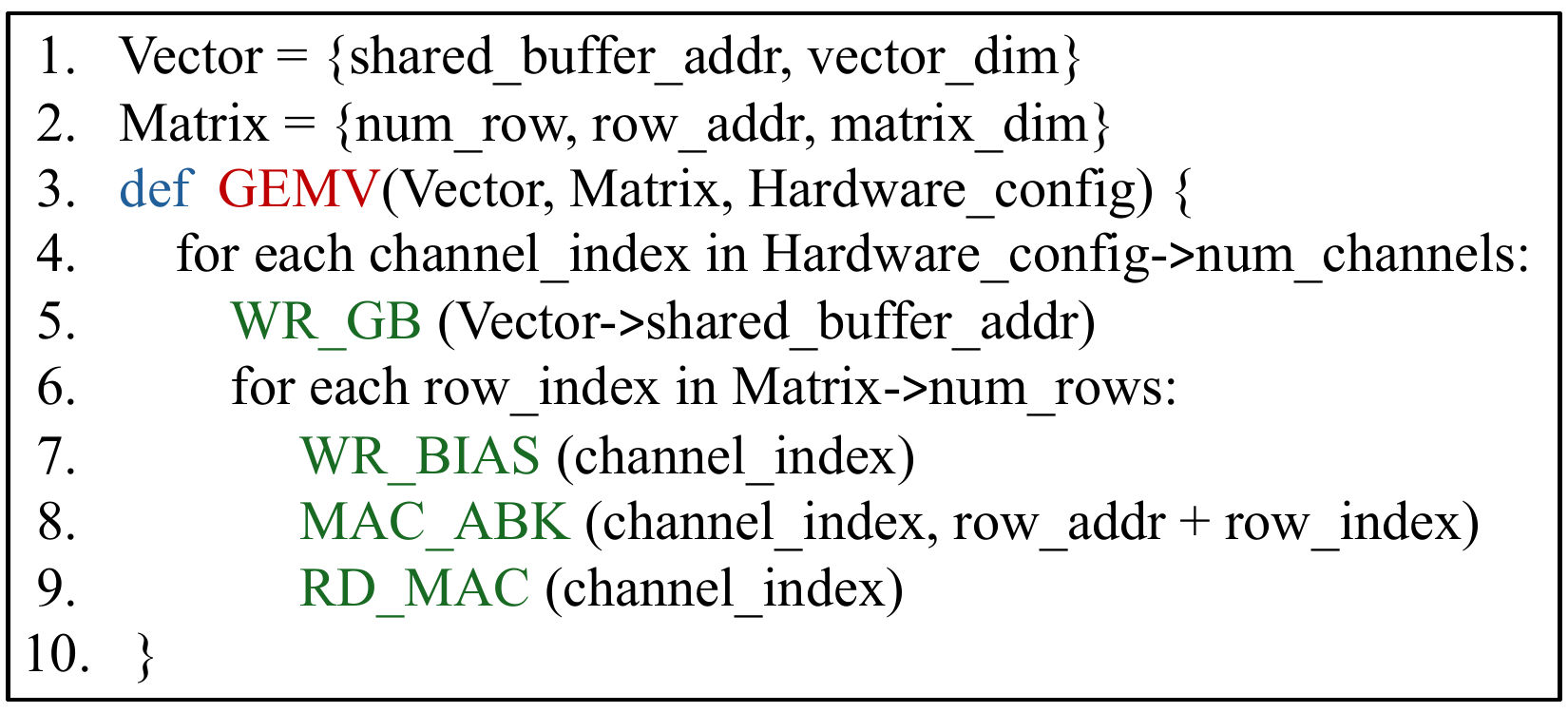}
    \caption{Vector-matrix multiplication compilation}
    \label{fig:Programming_model_code}
\end{figure}

Figure~\ref{fig:Programming_model_code} shows an example of compiling \texttt{GEMV} to \att{} instructions. Initially, the operands are designated to particular memory spaces, \textit{i.e.}, the vector operands in the \rf{} and the matrix operands in PIM channels (lines 1 and 2). \att{} instructions are then generated based on input operands' dimensions and memory addresses. Subsequently, the vector is copied to the Global Buffers in the PIM channels with \texttt{WR\_GB} instructions (line 5). This is followed by a sequence of operations for each matrix row within the near-bank PIM PUs. The \texttt{WR\_BIAS} instruction sets up the accumulation registers (line 7). \texttt{MAC\_ABK} performs the multiply-accumulate operations across all near-bank PUs in the PIM channel (line 8). Finally, \texttt{RD\_MAC} retrieves the results from the accumulation registers (line 9).

%% file: 5_methodology.tex
\section{Methodology}
\label{section:methodology}

Table~\ref{tab:Hardware configurations} lists the system configurations of \att{} and our GPU baseline.
The GPU system contains 4 NVIDIA A100 80GB GPUs equipped with the NVLink 3.0 interconnect.
\att{} has 32 CXL devices, resulting in a similar average power to the GPU system, as further explained in Section~\ref{subsec:power_results}. 

\begin{table}[h]
\footnotesize
\renewcommand\arraystretch{1.1}
\centering
    \caption{Evaluated system configurations}
    \label{tab:Hardware configurations}    
    \scalebox{1}{
        \begin{tabular}{|c||c|c|}
            \hline
            \textbf{System} & \textbf{\att{}} & \textbf{GPU} \\
            \hline
            \hline
            Hardware & 32 CXL devices & 4 NVIDIA A100 \\
            \hline
            Process & 1Y nm (14-16nm) & 7nm \\
            \hline
            Memory & 512GB, GDDR6 & 320GB, HBM2E \\
            \hline
            \multirow{2}{*}{\shortstack{Compute \\ Throughput}} & 512 TFLOPS (PIM) & \multirow{2}{*}{\shortstack{1248 TFLOPS}} \\
            \cline{2-2}
            & 96 TFLOPS (PNM) & \\
            \hline
            Peak Bandwidth & 512 TB/s (Internal) & 8 TB/s (External) \\
            \hline
            3-Year Owned TCO & 0.73\$/hour & 1.76\$/hour \\ 
            \hline
            3-Year Rental TCO & 1.05\$/hour & 5.45\$/hour \\ 
            \hline
            \hline
            GDDR6-PIM & \multicolumn{2}{|c|}{$t_{RCDRD}$=18ns, $t_{RAS}$=27ns, $t_{CL}$=25ns} \\
            Timing Constraints & \multicolumn{2}{|c|}{$t_{RCDWR}$=14ns, $t_{CCDS}$=1ns, $t_{RP}$=16ns} \\
            \hline
        \end{tabular}
    }
\end{table}

We benchmark Llama2 7B, 13B, and 70B models~\cite{touvron2023llama}.
Each evaluated query contains 512 tokens in the prefill stage and 3584 tokens in the decoding stage, adding up to a context length of 4K, \textit{i.e.,} the maximum supported by the Llama2 models.
For a fair comparison between \att{} and the GPU baseline, we deploy these models using different configurations for different parameter sizes: 1, 2, and 4 GPUs, and 8, 20 and 32 CXL devices.
We use vLLM~\cite{vLLM}, the \sota{} inference serving framework on GPUs with a batch size of 128, where the inference throughput saturates (Figure~\ref{fig:Context_Length}).

We generate \att{} instruction traces for a single block and verify the correctness using a functional simulator.
We modify Ramulator2~\cite{luo2023ramulator} to model a CXL device containing 32 GDDR6-PIM memory channels with timing constraints in Table~\ref{tab:Hardware configurations}. 
The inter-device communication through the CXL 3.0 protocol is modeled by an analytical model based on the CXL latency~\cite{li2023pond} and PCIe 6.0 bandwidth. 
To model a CXL switch supporting multicast, we use half of the bandwidth and double the latency of the baseline switch. 
We use Intel Xeon Gold 6430L CPU~\cite{intelxeon} as the host machine in \att{}.

We use Micron DRAM Power Calculator~\cite{micron-power-calculator} to evaluate DRAM core power using current and voltage specifications of Samsung's 8Gb GDDR6 SGRAM C-die~\cite{samsung-8gb-gddr6}.
The MAC operation power is modeled assuming 3$\times$ more current than a typical gapless read~\cite{aim2}.
We assume that each GDDR6 memory controller for two channels consumes 314.6 mW~\cite{dram-controller-power} and each BOOM RISC-V core consumes 250 mW~\cite{boom-pdf}.
We implement the RTL of the remaining components in the CXL controller and synthesize it using a TSMC 28nm technology library and the Synopsys Design Compiler~\cite{synopsis_dc}.
We find the critical path delay as 1ns at 28nm and project the CXL controller clock frequency to be 2.0 GHz at 7nm~\cite{scaling-technology}.

We estimate the die area of CXL controller in two parts. First, we synthesize the custom logic in 28nm~(See Table~\ref{tab:Area_and_power}) and scale it down to 7nm~\cite{scaling-technology}. Then, we add measurements of the memory controller, PCIe controller, and PHY from the NVIDIA GPU die shots~\cite{TU104, A100-die-shot}, which are also scaled down to 7nm. 
This results in an estimated area of 19.0$mm^2$ in 7nm.

\begin{table}[h]
\footnotesize
\centering
\caption{CXL Controller Custom Logic Area\&Power in 28nm}
    \label{tab:Area_and_power}    
    \scalebox{1}{
        \begin{tabular}{|c||c|c|c|}
            \hline
            \multicolumn{2}{|c|}{Components} & Area (mm$^2$) & Power (W) \\
            \hline
            \hline
            \multirow{2}{*}{SRAM} & Instruction Buffer & 3.33 & 0.61 \\
            \cline{2-4}
            & Shared Buffer & 0.11 & 0.03 \\
            \hline
            \multirow{3}{*}{Logics} & Accelerators & 1.34 & 0.18 \\
            \cline{2-4}
            & RISC-V Cores & 2.94 & 0.19 \\
            \cline{2-4}
            & Others & 0.12 & 0.05 \\
            \hline
            \hline
            \multicolumn{2}{|c|}{\textbf{Total}} & \textbf{7.85} & \textbf{1.06} \\
            \hline
        \end{tabular}
    }
\end{table}

\label{TCO}

Table~\ref{tab:Hardware configurations} presents the 3-year Total Cost of Ownership (TCO) for both owned and rental hardware. (a) \textit{Own TCO:} We model a local server by accounting for hardware and operational costs. (b) \textit{Rental TCO:} The cost for host CPU in \att{} and GPU are estimated based on the Microsoft Azure prices ~\cite{azure-price}. The CXL devices in \att{} are evaluated using the owned TCO methodology, as there are no available references for rental costs. To calculate operational cost, we use \$$0.139/KWh$~\cite{electricity-price} and average power consumption. Hardware costs are listed in Table~\ref{tab:hardware_cost}. While the lowest available price for A100 80GB is close to \$20,000, we instead use only \$10,000 by conservatively deducting 50\% margin~\cite{gpu-price}. The PIM module cost is estimated as 10$\times$ the cost of standard DRAM modules~\cite{pim-price, dram-price}.

\begin{table}[h]
    \footnotesize
    \centering
    \caption{Hardware Costs}
    \begin{tabular}{|c||c|c|}
        \hline
        \textbf{System} & \textbf{Hardware} & \textbf{Cost (\$)} \\
        \hline
        \hline
        \multirow{3}{*}{GPU} & Xeon Gold 6430 CPU~\cite{CPU-price} & 2,128 \\
        & 4 NVIDIA A100 80GB GPU~\cite{gpu-price} & 40,000 \\
        \cline{2-3}
        & \textbf{Total Cost} & \textbf{42,128}  \\
        \hline
        \hline
        \multirow{5}{*}{\shortstack{\att{} \\ 32 devices}} & Xeon Gold 6430 CPU~\cite{CPU-price} & 2,128 \\
        & 512GB GDDR6-PIM~\cite{pim-price, dram-price} & 11,873 \\
        & 32 CXL Controllers & 381.3 \\
        & 96-lane 48-port switch~\cite{switch-price} & 490 \\
        \cline{2-3}
        & \textbf{Total Cost} & \textbf{14,873} \\
        \hline
    \end{tabular}
    \label{tab:hardware_cost}
\end{table}

\begin{figure}[t]
    \centering
    \includegraphics[width=8cm]{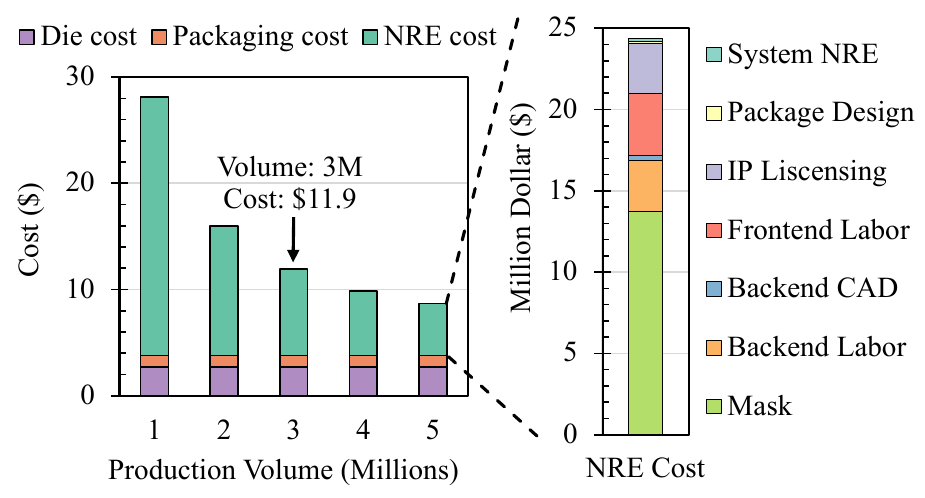}
    \caption{CXL Controller Cost Breakdown}
    \label{fig:TCO}
\end{figure}

Figure~\ref{fig:TCO} illustrates the breakdown of CXL controller cost per \att{} CXL device (Figure~\ref{fig:CXL_device}). The CXL controller costs are broken down into die, packaging and Non Recurring Engineering (NRE) cost components~\cite{ning2023supply, moonwalk}. Die cost is derived from the wafer cost, considering the CXL controller die area (19.0$mm^2$ in 7nm) and yield rate. A 300mm diameter 7nm wafer costs \$9,346 with a defect density of 0.0015 per $mm^2$~\cite{ning2023supply}. Cost of 2D packaging is assumed to be 29\% of chip cost~\cite{packaging-cost}, while the 2.5D packaging cost is calculated based on interposer, die placement and substrate assembly~\cite{palesko2014cost}.
NRE cost is influenced by chip production volumes, which we estimate at 3 million units based on the following assumptions.
NVIDIA shipped $3.76M$ datacenter GPUs in 2023~\cite{GPU-volume}.
We assume that $10\%$ of datacenter GPUs (around $370K$) are used for LLM inference.
Since each GPU consumes ${\sim}8\times$ more power compared to a CENT device (explained in Section~\ref{subsec:power_results}), we project ${\sim}3M$ volume for \att{} devices.

%% file: 6_results.tex
\begin{figure}[!b]
    \centering
    \includegraphics[width=\columnwidth]{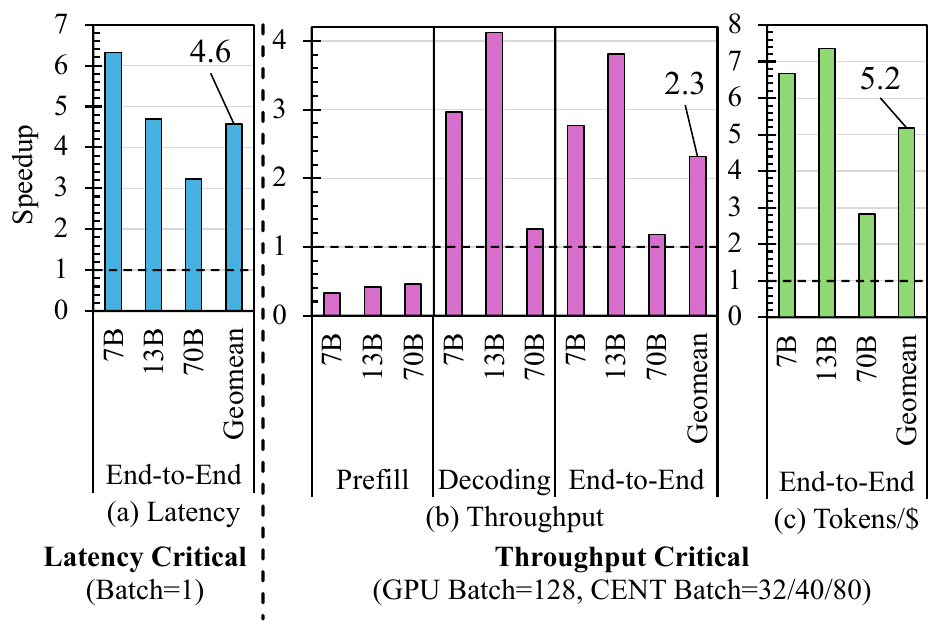}
    \caption{\att{} speedup over GPU baselines.}
    \label{fig:Main_results}
\end{figure}

\begin{figure*}[t]
	\centering
  	\includegraphics[width=17cm]{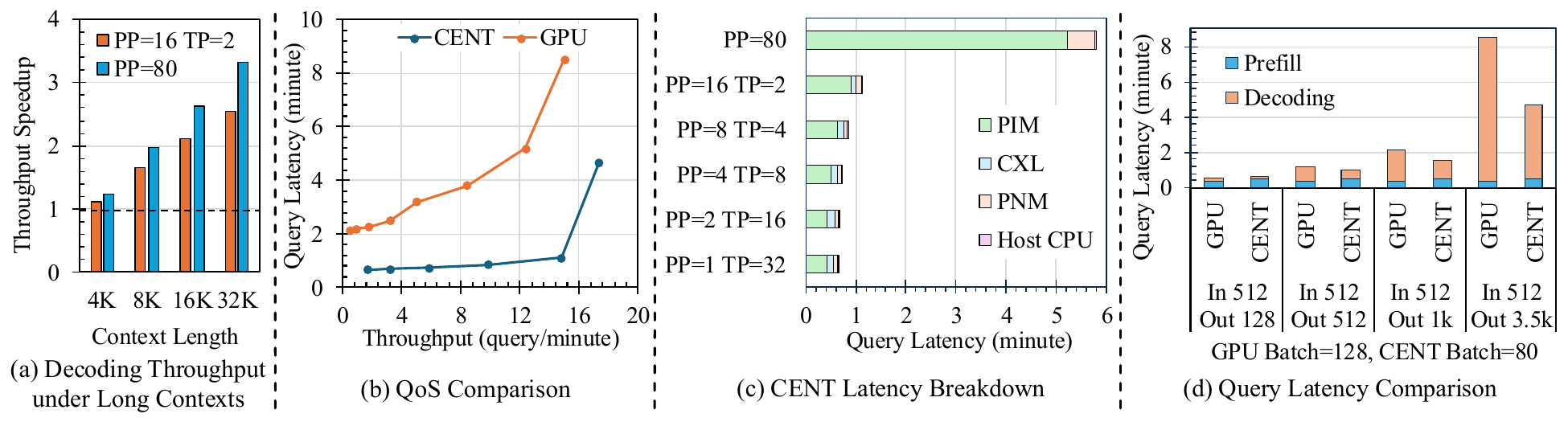}
	\caption{Analysis on Llama2-70B. 
 (a) \att{} achieves higher decoding throughputs with long context windows and 3584 decoding sizes (Section~\ref{section:methodology}). 16K and 32K context scenarios with PP=80 configurations require the 16Gb GDDR6-PIM module, increasing \att{} capacity to 1TB (2X of that used in main results). 
 (b) QoS analysis: \att{} provides lower query latency while achieving similar throughput as GPU. 
 (c) \att{} latency breakdown with different parallelism strategies.
 (d) Prefill (In) and decoding (Out) latency comparison with different In/Out sizes, at maximum supported batch size for both GPU and \att{}.
 }
\label{fig:Combo_Main_Lat_Breakdown}
\end{figure*}

\section{Results}

\subsection{\att{} versus GPU Baseline}~\label{subsec:main_results}

Figure~\ref{fig:Main_results} compares the performance of \att{} and our GPU baseline under two scenarios:
(a) \textit{Latency Critical:} We use a batch of 1 query (\att{}'s tensor parallel mapping).
In this case, \att{} reduces the end-to-end latency by 4.6$\times$ compared to GPUs.
This speedup is due to the higher internal memory bandwidth of PIM.
(b) \textit{Throughput Critical:} We use the maximum batch size of 128 for GPU experiments, as explained in Section~\ref{section:methodology}.
On the other hand, \att{} utilizes pipeline parallelism to enable batches of 32/40/80 queries for the three models (batch size = pipeline stages).
Using this configuration, \att{} achieves a geomean of 2.3$\times$ higher end-to-end throughput across three models.
\att{} demonstrates 1.2$\times$ speedup on Llama2-70B because this model applies the grouped-query attention technique~\cite{gqa}, improving the operational intensity of the attention layers.
Figure~\ref{fig:Main_results}(c) shows that \att{} processes 5.2$\times$ higher tokens per dollar than GPU, which attributes to \att{}'s higher throughput and 2.5$\times$ cheaper TCO (Table~\ref{tab:Hardware configurations}).

\begin{figure*}[t]
    \centering
    \includegraphics[width=17cm]{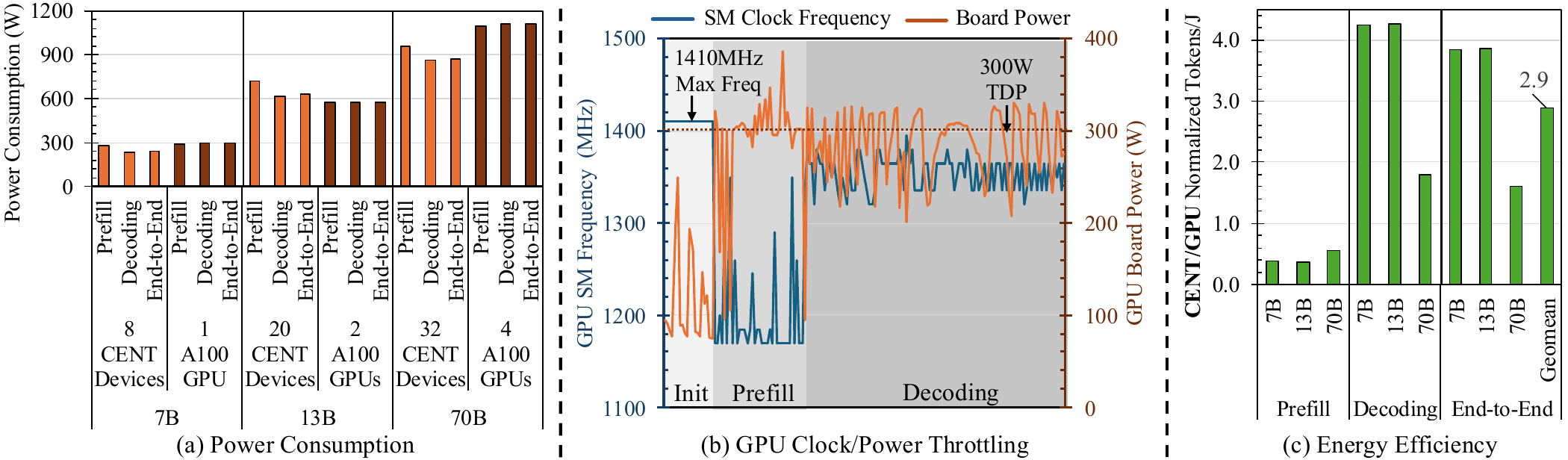}
    \caption{(a) Power consumption of \att{} and GPU (b) GPU SM frequency and board power, and (c) energy efficiency (Tokens per Joule) of \att{} and GPU
    using the maximum batch size, 512 prefill tokens and 3584 decoding tokens.}
	\label{fig:Energy_Power}
\end{figure*}

Figure~\ref{fig:Main_results}(b) compares throughput in the prefill and decoding stages.
GPU achieves 2.5$\times$ higher throughput in the compute-intensive prefill stage than \att{} due to GPU's 2.0$\times$ higher peak compute throughput.
Conversely, \att{} outperforms GPU in the memory-intensive decoding stage by 2.5$\times$ due to PIM's higher internal memory bandwidth.
Notably, the prefill stage accounts for only 2$\%$ of the total GPU end-to-end processing time, so the overall LLM inference throughput closely aligns with that of the decoding stage.


\begin{figure*}[t]
	\centering
  	\includegraphics[width=17cm]{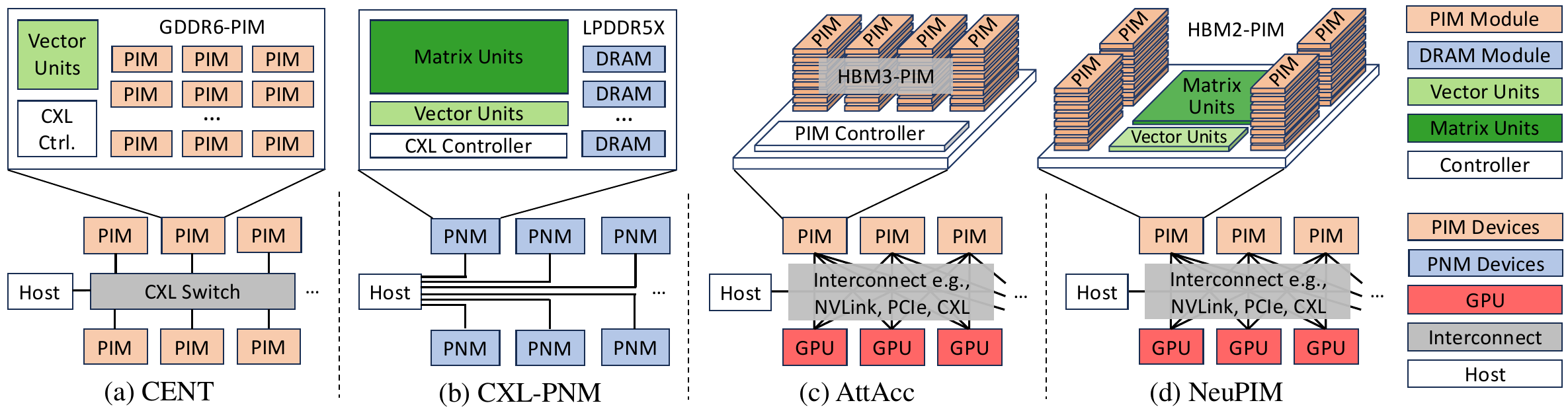}
	\caption{(a) \att{} employs vector units near PIM modules and utilizes a CXL switch to interconnect PIM devices with novel CXL communication primitives. (b) CXL-PNM~\cite{cxl-pnm} applies a processing-near-memory solution \emph{without} integrating compute logic into DRAM chips. (c-d) AttAcc~\cite{AttAcc} and NeuPIM~\cite{NeuPIM} are heterogeneous systems comprising GPUs and PIM devices.
    }
\label{fig:Comparison_related_works}
\end{figure*}

\textbf{\att{} performs better than GPU in long context scenarios.}
The results in Figure~\ref{fig:Main_results} use a 4K context window.
However, \sota{} LLMs support longer contexts, ranging from 128K to 1M tokens~\cite{gpt4-turbo, gemini-pro}.
As discussed in Section~\ref{GPU Performance Characterization}, with longer contexts, the GPU system saturates at smaller batch sizes, from batch=128 at 4K context to batch=16 at 32K context.
On Llama2-70B, \att{} achieves higher speedup than GPUs as context length increases, attaining up to 3.3$\times$ speedup in decoding throughput for a context length of 32K, as shown in Figure~\ref{fig:Combo_Main_Lat_Breakdown}(a). 

\textbf{\att{} has lower query latency than GPU at similar throughput.}
Figure~\ref{fig:Combo_Main_Lat_Breakdown}(b) illustrates our QoS comparison on Llama2-70B.
These results are collected with different batch sizes on GPUs and different TP/PP mapping strategies on \att{}.
\att{} provides 3.4-7.6$\times$ lower query latency while achieving similar throughput to the baseline GPU.

\textbf{Latency Breakdown.} Figure~\ref{fig:Combo_Main_Lat_Breakdown}(c) shows \att{}'s latency breakdown with different TP/PP mapping strategies.
PIM latency always dominates because most of the operations are mapped to PIM channels. As TP increases (from top to bottom), PIM latency reduces.
This is because more PIM channels are allocated to a single transformer block.
Yet, CXL communication latency increases with higher TP, because distributing a transformer block across more CXL devices necessitates more broadcast and gather transactions. 
Figure~\ref{fig:Combo_Main_Lat_Breakdown}(d) depicts the latency comparison between \att{} and GPU at maximum supported batch sizes.  
Compared to GPU, \att{} shows 1.4$\times$ higher latency in the prefill stage and 1.7-2.0$\times$ lower latency in the decoding stage. Decoding latency dominates the end-to-end latency.

\subsection{Power and Energy Consumption Analysis}\label{subsec:power_results}

We developed an \textit{activity-based} power model for \att{}. 
When deploying the Llama2-70B model on 32 CXL devices with the pipeline parallel model mapping, 27 devices are used. Among 80 transformer blocks (80 pipeline stages), 3 of them are mapped to each device, resulting in an average power of 32.4W per device. PIM operations and activation/precharge commands consume 54.5\% and 30.2\% of power, respectively.

Similarly, we used \texttt{nvidia-smi} to measure GPU power during the prefill and decoding stages in 100ms intervals. Figure~\ref{fig:Energy_Power}(a) illustrates the average power consumption of \att{} versus Nvidia A100 80GB GPUs. 
\textit{One} A100 GPU consumes $\approx8\times$ higher power than \textit{one} \att{} device. Modern GPUs consume significantly higher power as they support general-purpose PTX ISA~\cite{PTX-ISA}, a large number of Streaming Multiprocessors (108 SMs in A100), multithreading with fast context switching, and a multi-level cache hierarchy ($\approx$ 60 MB in A100~\cite{a100}). In contrast, \att{} is a custom architecture with minimal silicon used for near-bank compute units. 


GPUs operate near their thermal design power (TDP) of 300W~\cite{a100} during both the prefill and decoding stages when processing a large batch size of 128 queries.
Figure~\ref{fig:Energy_Power}(b) illustrates this by showing the GPU’s SM clock frequency and board power consumption for the Llama2-7B model.
During vLLM~\cite{vLLM} initialization, the clock frequency is maximized at 1410 MHz due to low compute throughput and memory bandwidth utilization.
In the prefill stage, high SM utilization signals the GPU’s power manager to throttle the clock frequency, maintaining power consumption within the TDP.
During the decoding stage, reduced SM utilization allows for an increase in clock frequency. A higher clock rate and memory bandwidth usage keep power near the TDP.

Figure~\ref{fig:Energy_Power}(c) shows that \att{} processes 2.9$\times$ more \emph{tokens per Joule} than GPU, on average.
In the compute-bound prefill stage, GPU is 2.4$\times$ more energy efficient, as it achieves efficient data reuse in the on-chip SRAM.
In the memory-bound decoding stage, \att{} achieves 3.2$\times$ higher energy efficiency, while GPU cannot efficiently reuse data in the SRAM because of the low operational intensity.
Our evaluation shows that \att{} consumes 0.6 pJ/bit on \texttt{MAC\_ABK} operations, making it $6.6\times$ more energy efficient than even \textit{only} the HBM2 memory read accesses of GPU, which consumes 3.97 pJ/bit~\cite{o2017fine}.

\subsection{\att{} versus PIM/PNM Baselines}

We compare \att{} with the \sota{} CXL-PNM~\cite{cxl-pnm} and heterogeneous GPU-PIM baselines~\cite{AttAcc, NeuPIM}. Figure~\ref{fig:Comparison_related_works} provides an architectural overview of these systems.

\begin{figure}[h]
	\centering
  	\includegraphics[width=8cm]{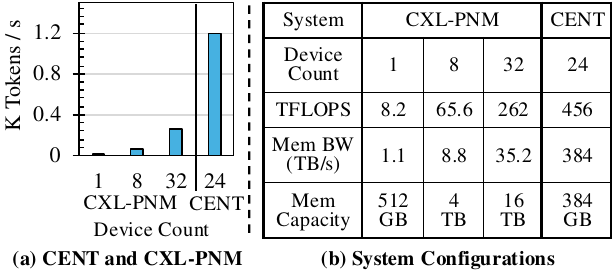}
    \caption{\att{} and CXL-PNM baseline comparison on OPT-66B~\cite{opt} with prefill=64 and decoding=1024.}
	\label{fig:CXL_PNM_OPT}
\end{figure}

\textbf{\att{} versus CXL-PNM.} Figure~\ref{fig:Comparison_related_works}(b) shows that CXL-PNM~\cite{cxl-pnm, samsung_pimpnm} is a processing-near-memory (PNM) platform that leverages a CXL controller to manage eight LPDDR5X packages within a single device. The CXL controller deploys matrix and vector units to perform computations \emph{near} commodity LPDDR5X chips. 
In contrast, Figure~\ref{fig:Comparison_related_works}(a) depicts \att{}, which utilizes processing-in-memory (PIM) technology to place compute logic adjacent to DRAM banks \emph{within} DRAM chips. Figure~\ref{fig:CXL_PNM_OPT}(b) shows that compared to CXL-PNM, \att{} provides significantly higher compute throughput (TFLOPs) and memory bandwidth (TB/s), at the cost of less memory capacity (GB).
Figure~\ref{fig:CXL_PNM_OPT}(a) illustrates that \att{}'s higher compute and memory bandwidth results in 4.5$\times$ higher throughput than CXL-PNM, at the maximum supported batch sizes for each system.

\textbf{\att{} versus GPU-PIM.} AttAcc~\cite{AttAcc} and NeuPIM~\cite{NeuPIM} are heterogeneous systems consisting of GPUs and PIM devices as shown in Figure~\ref{fig:Comparison_related_works}(c) and (d). The AttAcc system consists of 8 A100 GPUs with HBM3 memory~\cite{dgx-a100} and 8 HBM-PIM devices. Each HBM-PIM device consumes 116W and has a memory capacity of 80GB. The NeuPIM device integrates a TPUv4-like NPU~\cite{tpu} architecture near PIM modules and extends PIM with dual row buffers, enabling concurrent PIM-NPU memory access.  The evaluated NeuPIM platform comprises 8 A100 GPUs and 8 NeuPIM devices. 


Distinct from these systems, \att{} introduces a GPU-free inference server, providing an alternative cost-effective solution and eliminating the need for expensive GPUs. In GPU-PIM systems, the prefill stage is mapped to GPUs while the remaining computation is mapped to the PIM subsystem. \att{} does \emph{not} employ GPUs for the prefill stage for various reasons. \textit{First}, end-to-end LLM inference performance is primarily constrained by the decoding phase rather than the prefill phase; only 2\% of the total GPU's inference time is taken by the prefill stage across Llama2 models, on average (Section~\ref{subsec:main_results}). \textit{Second}, CENT’s compute throughput is not much worse than GPU ($\approx$49\%, Table~\ref{tab:Hardware configurations}). \textit{Third}, using expensive GPUs solely to support the prefill stage is a costly option. 
Using the methodology from Section~\ref{TCO}, we find that the TCO of AttAcc and NeuPIM is 3.5$\times$ and 2.6$\times$ higher than \att{}, respectively. The cost of HBM-PIM is estimated at 10$\times$ the price of HBM~\cite{HBM-price}, while the NPU cost is modeled based on die, 2.5D packaging, and NRE costs~\cite{tpu, palesko2014cost, moonwalk}.

\begin{figure}[t]
	\centering
  	\includegraphics[width=\columnwidth]{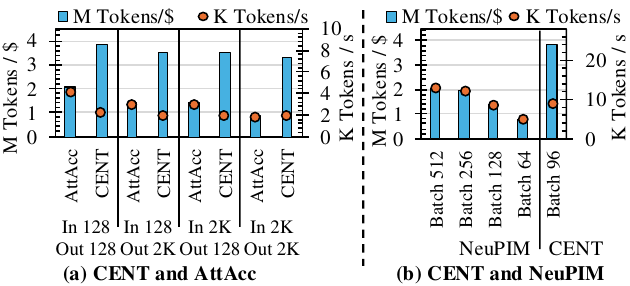}
	\caption{\att{} versus GPU-PIM (a) \att{} and AttAcc systems are evaluated on the GPT3-175B model across various input and output sizes, tested at the maximum supported batch sizes. 
 (b) The \att{} and NeuPIM systems are evaluated on GPT3-175B with data-parallel mapping (DP=4) and pipeline-parallel mapping (PP=4), respectively, using the ShareGPT dataset~\cite{sharegpt}. NeuPIM uses different batch sizes while \att{} uses the maximum supported batch size 96.}
\label{fig:AttAcc_NeuPIM}
\end{figure}

Figure~\ref{fig:AttAcc_NeuPIM} shows the performance of \att{} versus AttAcc and NeuPIM.
For a power-neutral evaluation, we assume 12 \att{} devices per GPU-PIM node. Across different sequence lengths and batch sizes, the blue bars show that \att{} processes 1.8-3.7$\times$ and 1.8-5.3$\times$ more tokens per dollar than AttAcc and NeuPIM systems, respectively. The orange dots show that \att{}'s raw throughput (Tokens/s) is 0.5-1.1$\times$ and 0.7-2.1$\times$ the throughput of AttAcc and NeuPIM, respectively. 
In scenarios with short sequence lengths, query batching enhances operational intensity in FC layers, improving performance on GPUs (or NPUs) with more TFLOPs. However, in cases with long sequence lengths that limit batch sizes, \att{} maintains higher raw throughput than the GPU-PIM baselines. Latest LLM models typically support 128K context windows~\cite{gpt4-turbo}. With these extended context lengths, we expect \att{} to provide even higher performance.

\subsection{Design Space Exploration}

\begin{figure}[h]
	\centering
  	\includegraphics[width=8cm]{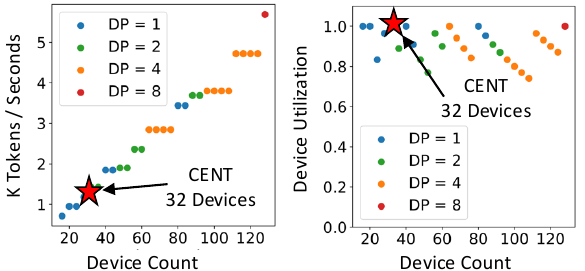}
    \caption{\att{} scalability study on Llama2-70B.}
	\label{fig:Scalability}
\end{figure}

\att{} can interconnect a flexible number of CXL devices, allowing for scalable system configurations. Figure~\ref{fig:Scalability} shows the scalability of \att{} on Llama2-70B from 16 to 128 devices, with throughput increasing from 0.68 K tokens/s to 5.7 K tokens/s. We start with pipeline-parallel (PP) mapping and then apply various levels of data-parallel (DP) mapping to further boost the throughput as the \att{} system scales up. 
As the number of devices increases, the throughput reaches intermittent plateaus at certain points. This is due to the inefficiency of distributing transformer blocks across CXL devices. For example, 80 transformer blocks in the Llama2-70B model can be allocated to 40 devices, with two blocks per device. Expanding from 40 to 44 devices results in a distribution of 1.8 blocks per device. Yet, dividing a single block across multiple CXL devices introduces substantial inter-device communication overhead, ultimately reducing performance. To mitigate this, we maintain the same block distribution with 44 devices as with 40, leaving the remaining 4 devices idle. 

The scalability of CXL devices is constrained by two primary factors: (1) The number of lanes and ports provided by a CXL switch. For example, a commercial PCIe 5.0 switch can accommodate up to 144 lanes and 72 ports~\cite{pcie5-switch}. (2) The maximum power supply available for the server, such as the DGX A100's peak input power of 6.5 kW~\cite{dgx-a100}. 
Due to these constraints, the \att{} system with a single switch can support up to 64 devices per server.
A larger number of devices can be driven by multi-socket CPUs or a memory pool implementation facilitated by two levels of CXL switches.

\subsection{Generality}

LLMs exhibit similar architectures but differ in their specific implementations of activation functions and positional encodings. \att{} is designed to support a variety of activation functions, including GeLU~\cite{GeLU}, Swish~\cite{Swish}, and their GLU variants~\cite{SwiGLU}. This versatility is achieved by decomposing these functions into fundamental non-linear operations, such as \texttt{sigmoid} and \texttt{tanh}, which are supported through lookup tables, as well as through basic PIM and RISC-V operations. Moreover, \att{} is capable of accommodating different types of positional embeddings, including both absolute~\cite{alibi} and relative~\cite{rope-paper} implementations. The integration of general-purpose RISC-V cores within the \att{} system opens up possibilities for further enhancements and optimizations of LLMs in the future.



%% file: 7_related_work.tex
\section{Related Work}

Various ML accelerators and HW/SW co-designs have recently been proposed~\cite{eyeriss, in-switch, maeri}. CXL memory expansion techniques are also widely explored~\cite{CXL-DL, CXL-DB, Demystify-CXL, cxl-memory-pool, jang2023cxl, gouk2022direct}. Sections~\ref{PIM} and ~\ref{PIM-prototype} already discuss PIM and PNM related works.

\noindent{\textbf{Transformer Accelerators.}} A variety of transformer accelerators~\cite{elsa, dota, sanger, FACT} have been developed to enhance this prevalent ML architecture.
TransPIM~\cite{transpim} accelerates inference of transformer encoders like BERT~\cite{bert} by reducing data loading time with an efficient token-based dataflow.
However, decoder-only LLM's inference tasks present a unique challenge due to their lower operational intensities, which have been less investigated. 
Approaches like Sprint~\cite{sprint}, OliVe~\cite{olive}, FABNet~\cite{FABNet}, and SpAtten~\cite{spatten} employ quantization, approximation, and pruning strategies, respectively, aimed at reducing computations within the transformer blocks, which are orthogonal to \att{}.

\noindent{\textbf{CXL-Based NDP Accelerators.}}  Samsung's CXL-PNM platform~\cite{cxl-pnm, samsung_pimpnm} integrates an LLM inference accelerator in the CXL controller. \att{} also integrates PIM memory chips with PUs adjacent to DRAM banks, providing both higher internal memory bandwidth and compute throughput than CXL-PNM.
Beacon~\cite{huangfu2022beacon} explores near-data processing in both DIMMs and CXL switches, with customized processing units for accelerating genome sequencing analysis.



%% file: 8_conclusion.tex
\section{Conclusion}

Given the challenges posed by the low operational intensity and substantial memory capacity requirements of decoder-only LLMs, we introduce \att{}, utilizing PIM technology to facilitate the high internal memory bandwidth and CXL memory expansion to ensure ample memory capacity. 
When compared to GPU baselines with the maximum supported batch sizes, \att{} achieves 2.3$\times$ higher throughput and consumes 2.3$\times$ less energy. \att{} also enables lower TCO and generates 5.2$\times$ more tokens per dollar than GPUs.


%% file: 9_artifact.tex
\appendix

\section{Artifact Appendix}

\subsection{Abstract}

This document provides a concise guide for reproducing the main performance, power, cost efficiency, and energy efficiency results of this paper in Figures~\ref{fig:TCO}, ~\ref{fig:Main_results}, ~\ref{fig:Combo_Main_Lat_Breakdown}, and ~\ref{fig:Energy_Power}.
The instructions cover the steps required to clone the GitHub repository, build the simulator, set up the necessary Python packages, execute the end-to-end simulation, process results, and generate figures.
The trace generator, performance simulator, power model, automation scripts, expected results, and detailed instructions are available in our \href{https://github.com/Yufeng98/CENT}{\red{GitHub repository}}.

\subsection{Artifact check-list (meta-information)}

{\small
\begin{itemize}
  \item {\bf Program:} C++ and Python.
  \item {\bf Compilation:} \texttt{g++-11/12/13} or \texttt{clang++-15}.
  \item {\bf Software:} \texttt{pandas}, \texttt{matplotlib}, \texttt{torch}, and \texttt{scipy} Python packages.
  \item {\bf Model:} Llama2 7B, 13B, and 70B~\cite{touvron2023llama}.
  \item {\bf Metrics:} latency, throughput (tokens/S), cost efficiency (tokens/\$), energy efficiency (tokens/J), and power.
  \item {\bf Output:} \href{https://github.com/Yufeng98/CENT/tree/main/figure_source_data}{\red{CSV}} and \href{https://github.com/Yufeng98/CENT/tree/main/figures}{\red{PDF}} files corresponding to Figures~\ref{fig:TCO}-\ref{fig:Energy_Power}.
  \item {\bf Experiments:} PIM trace generation and simulation, and \att{} power modeling.
  \item {\bf How much disk space is required?:} Approximately 100GB.
  \item {\bf How much time is needed?:} Approximately 24 hours on a desktop and 8~12 hours on a server.
  \item {\bf Publicly available?:} Available on \href{https://github.com/Yufeng98/CENT}{\red{GitHub}} and \href{https://zenodo.org/records/14776547}{\red{Zenodo}}.
  \item {\bf Code licenses:} \href{https://github.com/Yufeng98/CENT/blob/main/LICENSE}{\red{MIT License}}.
  \item {\bf Work automation?:} Automated by a few scripts.
\end{itemize}
}

\subsection{Description}

This artifact provides the necessary components to reproduce the main results presented in Figures~\ref{fig:TCO}, ~\ref{fig:Main_results}, ~\ref{fig:Combo_Main_Lat_Breakdown},  and ~\ref{fig:Energy_Power}.
It includes a trace generator, AiM simulator, power model, figure generator, and automation script.
While these figures incorporate simulation results from \att{}, they also rely on a baseline GPU system featuring four Nvidia A100 80GB GPUs, as detailed in Table~\ref{tab:Hardware configurations}.
Due to the high cost associated with these servers, only the expected results for the GPU baseline system are provided in the \href{https://github.com/Yufeng98/CENT/tree/main/data}{\red{data}} directory.

\subsubsection{How to access}

Clone the artifact from our GitHub repository using the following command. Please do not forget the \texttt{-}\texttt{-recursive} flag to ensure that the AiM simulator is also cloned:

\begin{lstlisting}
git clone --recursive https://github.com/Yufeng98/CENT.git
\end{lstlisting}


\subsubsection{Software dependencies}

AiM simulator requires \texttt{g++-11/12/13} or \texttt{clang++-15} for compilation.
The Python infrastructure requires \texttt{pandas}, \texttt{matplotlib}, \texttt{torch}, and \texttt{scipy} packages.


\subsubsection{Models}

Section~\ref{section:methodology} shows that we evaluate three Llama2 models~\cite{touvron2023llama}.
The model architecture and its PIM mapping are implemented in the \texttt{cent\_simulation/Llama.py} script.
The model weights are required only for the functional simulation of the PIM infrastructure. 
While the functional simulator is available in our GitHub repository, the performance simulator and power model described in this appendix do not model real values, as this does not impact the main results.
Consequently, the model weights and parameters are not required for this appendix.

\subsection{Installation}

\textbf{Building AiM Simulator.}
To build the simulator, use the following script:

\begin{lstlisting}
cd CENT/aim_simulator/
mkdir build && cd build && cmake ..
make -j4
\end{lstlisting}

\textbf{Setting up Python Packages.}
Install the aforementioned Python packages.
You can use the following script to create a \texttt{conda} environment:

\begin{lstlisting}
cd CENT/
conda create -n cent python=3.10 -y
conda activate cent
pip install -r requirements.txt
\end{lstlisting}

\subsection{Experiment workflow}

We provide scripts to facilitate the end-to-end reproduction of the results. The following steps outline the process.

\textbf{Generate and Simulate the Traces.}  
This step generates and simulates all required PIM traces.
It also processes the simulation logs, calculates individual latencies, and utilizes the \att{} power model to determine energy consumption and average power.
Upon completion, the generated trace and simulation log files will be stored in the \texttt{trace} directory, while the processed latency and power results can be found in \texttt{cent\_simulation/simulation\_results.csv}. 

\begin{lstlisting}
cd CENT/
bash remove_old_results.sh

cd cent_simulation/
bash simulation.sh [NUM_THREADS] [SEQ_GAP]
\end{lstlisting}

\textit{Note:} The argument \texttt{[NUM\_THREADS]} should be set according to the number of available parallel threads on your processor.
For instance, 8 threads are recommended for desktop processors, while server processors can utilize 96 threads. 

The argument \texttt{[SEQ\_GAP]} determines the gap between each simulated token.
Setting this value to one simulates every token sequentially, requiring approximately 100GB of disk space and taking around 24 hours on a processor with 8 threads or 12 hours on a processor with 96 threads.
To improve disk usage and reduce simulation time, the \texttt{[SEQ\_GAP]} argument can be set to a larger value, such as 128. This configuration simulates one out of every 128 tokens, processing token IDs of 128, 256, 384, and so on up to 4096.

\textbf{Process the Results.}  
This step processes the simulation results and computes the latency, throughput, power, and energy for the prefill, decoding, and end-to-end phases.
After processing the results, this script stores them in this file: \texttt{cent\_simulation/processed\_results.csv}.

\begin{lstlisting}
cd CENT/cent_simulation/
bash process_results.sh
\end{lstlisting}

\textbf{Generate Figures.}  
The following script generates Figures~\ref{fig:TCO}-\ref{fig:Energy_Power}.
This process utilizes the baseline GPU results, available in the \href{https://github.com/Yufeng98/CENT/tree/main/data}{\red{data}} directory, along with the processed results.
It computes the normalized results and generates both a PDF file containing the figures and a CSV file with the corresponding numerical data.

\begin{lstlisting}
cd CENT/
bash generate_figures.sh
\end{lstlisting}

\subsection{Evaluation and expected results}

The normalized results and the figures will be located in the \texttt{figure\_source\_data} and \texttt{figures} directories.
The expected results can be found in Figures~\ref{fig:TCO}-~\ref{fig:Energy_Power} or in the generated \href{https://github.com/Yufeng98/CENT/tree/main/figure_source_data}{\red{CSV}} and \href{https://github.com/Yufeng98/CENT/tree/main/figures}{\red{PDF}} files on our GitHub repository. Figures in the paper are generated using Microsoft Excel. To visualize the figures in the paper's format, copy the normalized data from the CSV files to the \texttt{Data} sheet of the provided \href{https://github.com/Yufeng98/CENT/blob/main/cent_simulation/Figures.xlsx}{\red{Figures.xlsx}}.
Figures will be generated in the \texttt{Figures} sheet.

\clearpage




